\documentclass[conference,a4paper]{IEEEtran}
\IEEEoverridecommandlockouts
\usepackage{cite}
\usepackage{amsmath,amssymb,amsfonts}
\usepackage{algorithmic}
\usepackage{graphicx}
\usepackage{textcomp}
\usepackage{tikz}
\usepackage{xcolor}
\usepackage{algorithm}
\usepackage{subcaption}
\usepackage{soul}
\usepackage{booktabs}
\usepackage{multirow}
\usepackage{adjustbox}
\usepackage{enumitem}
\usepackage{hyperref}
\usetikzlibrary{plotmarks}

\definecolor{tabblue}{rgb}{0.12156863, 0.46666667, 0.70588235}
\definecolor{taborange}{rgb}{1.0, 0.49803922, 0.05490196}
\definecolor{tabgreen}{rgb}{0.17254902, 0.62745098, 0.17254902}
\definecolor{tabred}{rgb}{0.83921569, 0.15294118, 0.15686275}
\definecolor{tabpurple}{rgb}{0.58039216, 0.40392157, 0.74117647}
\definecolor{tabbrown}{rgb}{0.54901961, 0.3372549, 0.29411765}
\definecolor{tabpink}{rgb}{0.89019608, 0.46666667, 0.76078431}
\definecolor{tabgray}{rgb}{0.49803922, 0.49803922, 0.49803922}
\definecolor{tabolive}{rgb}{0.7372549, 0.74117647, 0.13333333}
\definecolor{tabcyan}{rgb}{0.09019608, 0.74509804, 0.81176471}

\def\BibTeX{{\rm B\kern-.05em{\sc i\kern-.025em b}\kern-.08em
    T\kern-.1667em\lower.7ex\hbox{E}\kern-.125emX}}
\begin{document}

\title{Routing in Quantum Networks with End-to-End Knowledge\\
}

\author{
    \IEEEauthorblockN{
        Vinay Kumar\IEEEauthorrefmark{1}\IEEEauthorrefmark{2}\IEEEauthorrefmark{3}\thanks{\IEEEauthorrefmark{3}Email: vinay.kumar@phd.unipi.it}, 
        Claudio Cicconetti\IEEEauthorrefmark{2}\IEEEauthorrefmark{4}\thanks{\IEEEauthorrefmark{4}Email: c.cicconetti@iit.cnr.it},
        Marco Conti\IEEEauthorrefmark{2}\IEEEauthorrefmark{5}\thanks{\IEEEauthorrefmark{5}Email: marco.conti@iit.cnr.it}, and
        Andrea Passarella\IEEEauthorrefmark{2}\IEEEauthorrefmark{6}\thanks{\IEEEauthorrefmark{6}Email: a.passarella@iit.cnr.it}
    } \\
    \IEEEauthorblockA{\IEEEauthorrefmark{1}Department of Information Engineering, University of Pisa, Pisa, Italy} \\
    \IEEEauthorblockA{\IEEEauthorrefmark{2}Institute for Informatics and Telematics (IIT), National Research Council (CNR), Pisa, Italy}
}

\maketitle

\begin{abstract}
Given the diverse array of physical systems available for quantum computing and the absence of a well-defined quantum internet protocol stack, the design and optimisation of quantum networking protocols remain largely unexplored. To address this, we introduce an approach that facilitates the establishment of paths capable of delivering end-to-end fidelity above a specified threshold, without requiring detailed knowledge of the quantum network's properties. In this study, we define algorithms that are specific instances of this approach and evaluate them in comparison to Dijkstra's shortest path algorithm and a fully knowledge-aware algorithm through simulations. Our results demonstrate that one of the proposed algorithms consistently outperforms the other methods in delivering paths above the fidelity threshold, across various network topologies and the number of source-destination pairs involved, while maintaining significant levels of fairness among the users and being robust to inaccurate estimations of the expected end-to-end fidelity.
\end{abstract}


\bigskip
\noindent\textbf{Note.} This is the author accepted manuscript. \\
\textbf{Please cite as}: Kumar, V., et al.: Routing in quantum networks with end‐to‐end knowledge. 
\textit{IET Quantum Communication}, e70000 (2025). 
\href{https://doi.org/10.1049/qtc2.70000}{https://doi.org/10.1049/qtc2.70000}


\section{Introduction}\label{sec:introduction}
The rapid advancements in quantum computing continue to enhance our understanding of applications that are either beyond the reach of classical computing or difficult to achieve for it. These applications include quantum cryptography \cite{Bennett14}, drug discovery \cite{aspuru05, cao19}, optimisation problems \cite{farhi14}, quantum machine learning \cite{biamonte17}, and quantum simulations \cite{georgescu14}. In response to the looming threat posed by the potential breaking of many current cryptographic systems — through the efficient solving of problems like integer factorisation, which underpins RSA encryption, using Shor's algorithm \cite{shor94} — the emergence of quantum networks offers a promising solution \cite{kimble08, elliott02, wehner18}. These networks have the potential to revolutionise communication by harnessing the power of fundamental quantum phenomena.

To unlock applications such as blind quantum computing \cite{broadbent09}, distributed quantum computing \cite{cacciapuoti19, caleffi18}, quantum key distribution \cite{ekert91, pirandola20, singal22}, and clock synchronisation \cite{komar14}, quantum networks must be capable of providing entangled states among users. A quantum repeater, responsible for the distribution of entanglement among users, is a critical component in quantum networks. Quantum repeaters can vary significantly, not only in the physical realisation of their qubits but also in the nature of their connections \cite{briegel98, duan01, childress06, simon07, zhao07, sangouard09, sangouard11, li16, Perseguers08, jiang09, munro12, muralidharan14}. Different schemes for quantum repeaters have been proposed depending on the error correction techniques employed, which are crucial for long-distance communication \cite{munro15, muralidharan16}.

Similar to how the classical internet uses the Open Systems Interconnection (OSI) model to divide network functionalities and thus enhance scalability and simplicity, a quantum internet protocol stack can be defined. Different research groups have proposed several such models based on varying functionalities of layers, approaches to error correction, and assumptions about entanglement generation \cite{illiano22}. Despite the lack of a commonly accepted quantum internet protocol stack, the current literature allows us to categorise quantum internet protocols into physical, link, network, transport, and application layer protocols \cite{yuan24}.

This work is aimed at developing network layer protocols concerning bipartite entanglement routing, which essentially entails the selection of a path to establish end-to-end entanglement between two remote nodes in the network, with a minimum level of required \emph{fidelity} (see Section~\ref{sec:net_and_mot} for a precise definition of fidelity in our context).
This problem has already been studied in the literature for different target scenarios and objectives (see Section~\ref{sec:works}), but a common trait to all the previous works is that they assume some knowledge about the quantum repeaters and the links interconnecting them, which must be sufficiently detailed to determine the best path according to some composed metric.
However, it is not yet clear whether the future protocols defined for the quantum internet will make this knowledge readily available to the entity making decisions on path selection.
To fill this gap, in this work, we relax the assumption on detailed per-node/per-link knowledge and study instead the problem of path selection in a quantum network with end-to-end knowledge; we call this approach ``grey-box'', as opposed to ``white-box'' commonly adopted in the state of the art.
The major contributions of this work are listed below:
\begin{itemize}
    \item We introduce a novel robust grey box approach to quantum network routing that does not require detailed information about the network nodes and links, but relies on an end-to-end characterisation of the quantum network interconnecting two specific endpoints, motivated by recent proposals in this area, such as~\cite{helsen23}.
    \item We present a comprehensive comparison between the proposed approach and an alternative that requires full per node/link knowledge. We consider random (according to the well-known Waxman model~\cite{wax88}) and regular topologies (square grid lattice), and we study the performance at various levels of resources availability in the network, namely the number of nodes and links.
    \item We demonstrate that the proposed $kx_{0}$ algorithm (\ref{ssec:routing}) consistently outperforms other methods in achieving higher end-to-end fidelity across various network topologies and source-destination pairs employed while maintaining significant levels of fairness among users.
\end{itemize}
The organisation of the paper is as follows. In Section~\ref{sec:works}, we discuss the state of the art. In Section~\ref{sec:net_and_mot}, we formulate the system model and the routing problem. In Section~\ref{sec:sim_and_method}, we present the simulation methodology. In Section~\ref{sec:results}, we present and analyse the results. Finally, in Section~\ref{sec:conclusions}, we conclude the paper and discuss some possible future research directions. 

\section{State of the art}\label{sec:works}
While the quantum routing problem has been extensively studied in the literature, finding an optimal path that considers all aspects of the problem remains an open challenge. An ideal routing policy should enable end-to-end entanglement between endpoints with maximum throughput, ensuring maximum fidelity or fidelity above a specified threshold, minimising resource consumption and latency, and ensuring fairness among users.
Additionally, the complexities arising from the quantum network structure itself, if considered in a realistic scenario, include heterogeneous links and quantum repeaters, EPR generation rate, EPR success rate, and purification success rate. 

Among the initial investigations, Meter et al. \cite{meter13} proposed the inverse of the throughput of the link as the routing metric which is Bell pairs per second at a certain fidelity. The other possible candidates for routing metrics such as the total number of laser pulses or the total number of measurement operations, were found to be unsuitable. The proposed metric (inverse of throughput) takes into account the quality of links which arise from their length. Caleffi et al. \cite{caleffi17} put forward a routing protocol which used end-to-end entanglement rate of the path as the routing metric. This metric takes into account physical mechanisms such as entanglement generation, swapping, Bell state measurement, and decoherence time. 
Pant et al. \cite{pant19} and Li et al. \cite{li21} proposed routing schemes that simply utilise the hop count as the routing metric, which we use as a baseline reference in our performance evaluation in Section~\ref{sec:results}.

Shi et al. \cite{shi20} introduced the QCAST framework, which
adopts a time-slot structure, a modified version of the approach proposed by Pant et al. \cite{pant19}. Each time-slot consists of four phases. During phase one, all nodes receive information regarding the current source-destination pair intending to establish end-to-end entanglement. In phase two, routes are determined for the requests from phase one using the QCAST algorithm, and external entanglement links are established between neighbouring nodes, utilising the qubits and channels associated with these nodes. Phase three involves nodes exchanging link state information through classical channels. Finally, in phase four, each node establishes internal links to complete the route between the source and destination. That work has inspired us to define the system, illustrated in Section~\ref{sec:net_and_mot}, which also involves multiple phases in a time-slotted system. Zhang et al. \cite{zhang21} introduced a novel fragmentation-aware algorithm designed to address resource contention issues across multiple paths. The core concept involves redefining network resources as `public' (shared among all paths) rather than `private' (dedicated to specific paths), contrasting with the approach taken by QCAST. Zhao et al. \cite{zhao21} introduced a method aimed at enhancing network throughput over QCAST by as much as 68.75\%, achieved through the provision of redundant external links established during phase two. Such optimisations are orthogonal to the problem addressed in this work and, in principle, they can be used in combination with the solutions proposed.

Despite its importance for distributed quantum computing applications, fidelity is often overlooked in the literature as a routing metric. Li et al. \cite {li22} proposed an iterative routing algorithm based on entanglement purification to provide fidelity-guaranteed paths. In \cite{vk23} we proposed a vertex-weighted routing algorithm that delivers paths above certain threshold fidelity for heterogeneous nodes with mixed efficiency figures. Other routing metrics include latency \cite{cicconetti21, zhang22}, fairness \cite{yang22}, and number of repeaters \cite{rabbie22}.

A common trait of above all routing metrics is that they require detailed knowledge of the network characteristics with certainty such as link quality or length \cite{meter13}, EPR generation rate \cite{caleffi17}, EPR success rate \cite{shi20, zhang21, zhao21}, purification success rate \cite{li22}, and node quality \cite{vk23}. We classify approaches that construct routing metrics based on detailed network information as ``white-box" approaches.
In this work, we relax such an assumption and design a solution for path selection that only uses knowledge of the network topology, i.e., how the quantum repeaters are interconnected via quantum links, and an estimate of the end-to-end fidelity between end nodes. We term this method a ``grey-box" approach, as it does not demand precise network information but relies on select, relevant insights.

\section{Network and Motivation}\label{sec:net_and_mot}
In this section, we establish the network design and motivation for this work. In Section~\ref{ssec:architecture}, we provide a comprehensive discussion of the model of the system. In Section~\ref{ssec:estimations}, we discuss end-to-end fidelity estimation in the grey-box approach. In Section~\ref{ssec:end-to-end_entanglement}, we discuss end-to-end entanglement. In Section~\ref{ssec:slotted_model}, we discuss the slotted model of the time evolution. In Section~\ref{ssec:routing}, we establish the problem of routing in quantum networks.

\subsection{System Model}\label{ssec:architecture}
\begin{figure}[tb]
    \includegraphics[width=\columnwidth]{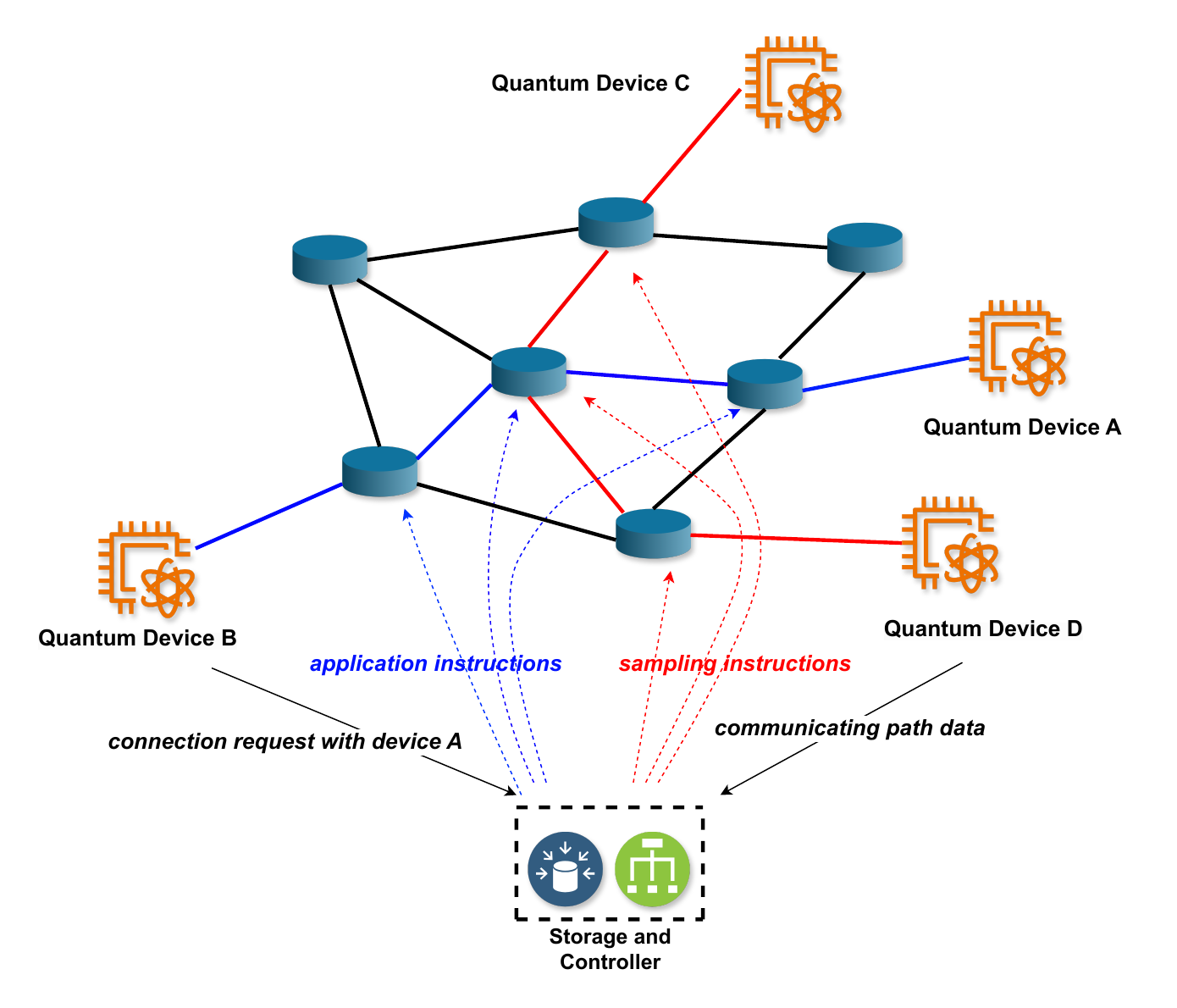}
    \vspace{-0.6cm}
    \caption{Quantum network architecture}
    \label{fig:network_architecture}
\end{figure}

In this paper we consider the general quantum network structure shown in Figure~\ref{fig:network_architecture}. It consists of the following key components:
\begin{enumerate}[label=\roman*.]
    \item \textbf{Quantum Repeater:} Fundamental component in quantum networks, which allows for long-distance end-to-end entanglement.
    \item \textbf{Quantum Device:} Quantum computer capable of running quantum applications connected to the quantum network.
    \item \textbf{Storage and Controller:} A logically centralised entity that maintains the network topology and an estimate of the end-to-end fidelity between pairs of quantum devices (storage) and is in charge of making routing decisions to establish end-to-end entanglement between them. 
    \item \textbf{Quantum Network Link:} A link with quantum and classical channels between two neighbour quantum repeaters or a quantum repeater and a quantum device.
\end{enumerate}
We assume that for every link there is a heralded source of entangled photons which are transmitted to each end of the link and enter a photon detector. Therefore, each node (quantum device or repeater) is assumed to be equipped with one photon detector for each of its links.  Furthermore, the quantum repeaters are required to have a sufficient number of Bell state analysers that can be used to perform entanglement swapping, as illustrated below. Finally, we assume for simplicity that no time multiplexing is performed: for as long as two links of a quantum repeater are assigned to a specific end-to-end path, they cannot be used by other paths.
Two quantum devices, e.g., A and B, can make use of the quantum network to realise distributed quantum applications through end-to-end entanglement of (some of) their qubits.
We assume that this happens through a request to the controller, which is in charge of deciding the path between A and B among all the possible ones in the network.
Once such a path is selected, the controller is expected to notify all the quantum repeaters involved so that they perform appropriate operations to realise the end-to-end entanglement between quantum devices A and B until further notice.
Such notifications are referred to as ``application instructions'' in Figure~\ref{fig:network_architecture}, and we assume here,
for simplicity, that a link can be assigned a single path, i.e., it cannot be shared across applications between different pairs of quantum devices.
In the literature, as reviewed in Section~\ref{sec:works}, it is common to assume that the controller will make use of detailed knowledge about all the quantum repeaters to select a path in a way that goes in the direction of maximising some performance objective, such as throughput or quality of entanglement.
In particular, it is typically assumed that the controller knows:

\begin{enumerate}
    \item The topology of the network, i.e., it knows which quantum repeaters share a link.
    \item Quantum repeater-specific information that can be used to estimate the nominal performance of end-to-end entanglement, usually measured through the \emph{fidelity}, for a given path.
\end{enumerate}

\subsection{End-to-end fidelity estimation}\label{ssec:estimations}
As introduced in Section~\ref{sec:introduction}, in this work we question the latter assumption: estimating the fidelity of an end-to-end entanglement in real conditions would be an overly complex task since the latter is determined by a complex combination of time- and technology-dependent processes involving, among the others, the heralding of local entanglement on a link \cite{hang21}, the purification of such an entanglement \cite{chen24}, the decoherence of quantum memories \cite{pompili21}, the link-level protocols \cite{axel19}.
We therefore advocate an alternative approach, in which such end-to-end fidelity is not estimated \emph{a~priori} but measured \emph{a~posteriori} through an in-band mechanism running in the network, which, e.g., assumes periodic activation of ``sampling instructions'' (Figure~\ref{fig:network_architecture}) on unused links for the purpose of acquiring/refreshing the information on the measured fidelity between any two quantum devices in the network.
For instance, randomised benchmarking techniques could be used for this~\cite{helsen23}.
The specific procedure that can be used for this purpose is beyond the scope of this work, in which we merely assume that the so-called Storage component is made aware of the (average) fidelity that can be achieved for any given path, without knowing the details of how much each link/node contributes to the end-to-end result.

We are aware that the number of possible paths in a generic network can be very large, which would render this approach practically unfeasible, but we make the following observation to advocate such a solution, at least in some scenarios. The topology of a network is expected to remain stable for a time that is much longer than the time scales relevant for execution of distributed quantum computing applications, as modifying the topology implies the installation of a device (quantum computer or repeater) and its physical interconnection with the quantum network or provisioning additional infrastructures (satellite or optical fibre).
Therefore, the information about end-to-end fidelity may be accumulated in the Storage over time through a background process, prioritising shorter paths between pairs of end nodes, which consume fewer resources and are more likely to exhibit a higher fidelity. However, it is crucial to consider that network topology information can be influenced by environmental factors beyond physical layout changes. Such environmental interactions may alter link performance, thereby impacting the accuracy of end-to-end fidelity predictions.

Consequently, developing efficient methods to manage the accumulation and updating of end-to-end fidelity knowledge remains an essential area for further research, as noted in recent studies (e.g., [35]). This topic, while outside the scope of this work, is complementary to our approach. In the meantime, given the early stages of development in this direction, we have implemented robustness tests to evaluate our methods under conditions with unreliable or outdated end-to-end knowledge, as will be discussed in subsequent sections.
\begin{figure}[tb]
    \includegraphics[width=\columnwidth]{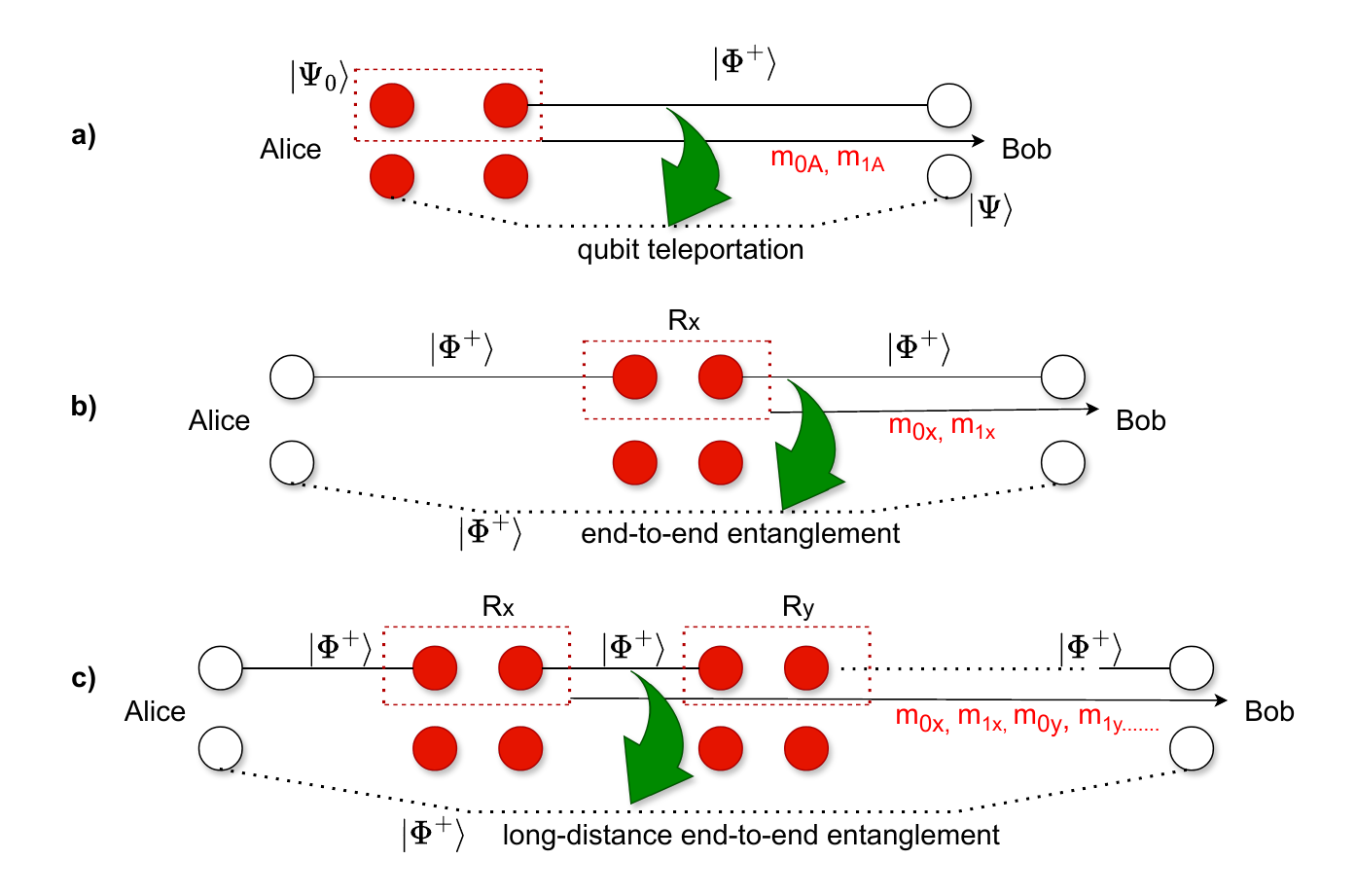}
    \vspace{-0.6cm}
    \caption{\textbf{a) Teleportation:} A bell state measurement by Alice transports initial state $| \Psi_0 \rangle$ to Bob as $| \Psi \rangle$. \textbf{b) Entanglement Swapping:} A bell state measurement at quantum repeater $R_x$ swaps two entangled pairs ($| \Phi^{+} \rangle$) into end-to-end entanglement between Alice and Bob. \textbf{c) A linear network:} A bell state measurements at quantum repeaters $R_i$ for $i \in \{x, y, ...\}$ swaps $i+1$ entangled pairs ($| \Phi^{+} \rangle$) into a long-distance end-to-end entanglement between Alice and Bob.}
    \label{fig:teleport_swap_chain}
\end{figure}
\subsection{End-to-end entanglement}\label{ssec:end-to-end_entanglement}
We now delve into the details of the establishment of end-to-end entanglement.
As exemplified in Figure~\ref{fig:network_architecture} end-to-end entanglement between quantum devices C and D or quantum devices B and A is carried out via entanglement swapping as shown in Figure~\ref{fig:teleport_swap_chain}c). The process is as follows:
\begin{enumerate}[label=\roman*.]
    \item All the EPR pairs between neighbouring quantum repeaters along the intended route are successfully shared via the respective quantum links.
    \item The quantum repeaters along the route perform a Bell state measurement on one qubit of the shared EPR pair.
    \item They send the classical results of Bell measurement to one of the end nodes (e.g., Bob) and inform the other one (Alice) about the procedure's success.
    \item Bob applies Pauli gates ($I$, $\sigma_x$, $\sigma_z$ or $\sigma_x \sigma_z$) depending upon the classical results of measurement received.
\end{enumerate}
Bell state measurement performed on each quantum repeater is central to establishing end-to-end entanglement and it relies on the ability to project onto the basis states, which can be summarised through an $\eta$ parameter as follows \cite{helstrom69, dur99}:
\begin{equation}\label{eq:P0}
P_0 = \eta |0 \rangle \langle 0| + (1 - \eta) |1 \rangle \langle 1|
\end{equation}
\begin{equation}\label{eq:P1}
P_1 = \eta |1 \rangle \langle 1| + (1 - \eta) |0 \rangle \langle 0|
\end{equation}
For $\eta = 1$, the device always measures the qubit on the correct basis. However, for $\eta < 1$ the device measures the qubit on the correct basis with an expectation value of $1 - \eta$.
In this study, we consider the efficiency of the measurement devices integral to the quantum repeater's overall performance. Consequently, the errors introduced by the Bell state measurements influence the categorisation of quantum repeaters i.e., a quantum repeater belongs to a certain class $g$ depending upon the efficiency figure $\eta_g$ of its measurement devices, which, in turn, affects the fidelity of the end-to-end entangled pairs used by the applications running in the quantum devices.
In fact, for a linear route connecting two quantum devices through $\sum_i^G N_g$ quantum repeaters belonging to $G$ classes, each class containing $N_g$ quantum repeaters, the fidelity is given by \cite{dur99}:
\begin{equation}\label{eq:fidelity}
\begin{split}
F_p = \frac{1}{4} \biggl\{ 1+3 \prod_{g=1}^G \left[ \left( \frac{4 {\eta_g}^2 -1}{3} \right)^{N_g} \left( \frac{4 F -1}{3} \right)^{N_g} \right] \\
\left[\frac{4
F -1}{3}\right] \biggr\},
\end{split}
\end{equation}
where each pair of qubits in a link is assumed to be in a Werner state with initial fidelity $F$ upon generation of the local entanglement. It is worth noting that this equation models the heterogeneity across quantum repeaters, where each class of quantum repeater has a distinct efficiency figure \( \eta_g \). The end-to-end fidelity \( F_p \) thus becomes a function of both the number of repeaters in each class and their respective efficiency figures. This heterogeneity influences the cumulative fidelity of entanglement over the entire route, thereby impacting the performance of algorithms designed to optimize routing paths in quantum networks. In addition, the initial fidelity of EPR pairs in quantum links can vary, leading to heterogeneity in quantum links that impacts the end-to-end fidelity calculation. This end-to-end fidelity can be directly utilized in the grey-box algorithms to guide path selection and optimization. However, we assume uniform initial fidelity of EPR pairs $F$ across all quantum links in our analysis for simplicity.


The general dynamics of delivering entanglement between sources and destinations in a quantum network are depicted in Figure~\ref{fig:time}.
We assume that the network topology, modelled as an undirected graph $G(V, E)$, is static and known to the controller.
Such an assumption is reasonable for a local or metropolitan quantum network that is owned or operated by the same entity and it is, in fact, valid for intra-domain routing protocols in classical networks, such as Open Shortest Path First (OSPF) and more recent solutions relying on Software Defined Network (SDN) concepts.

At the time of composing this paper, Liu et al. \cite{Liu24} have recently published research that partially overlaps with the themes of our study. They introduced an approach for inter-domain routing among Autonomous Systems (ASs), utilising a methodology that involves selecting the best $k$ paths based on a randomised benchmarking procedure. This procedure estimates the end-to-end quality of each path \cite{helsen23}. Despite the apparent similarities, there are critical distinctions to be underscored.
Primarily, our research focuses on intra-domain quantum routing, thus addressing a fundamentally different aspect of network as compared to the inter-domain routing proposed by Liu et al. \cite{Liu24}. Although our approach also prioritises the selection of the best $k$ paths, our criteria hinge on the underlying network topology, which dictates the probability mass function (PMF) of path lengths and their corresponding end-to-end fidelities, which is integral to solving intra-domain routing challenges. Unlike the inter-domain framework proposed by Liu et al., which manages connectivity across multiple networks, our work is specifically designed for single-network, intra-domain scenarios where the routing scheme leverages detailed internal network knowledge rather than external network relationships.
Moreover, we advance a more granular analysis of path fidelity by beginning with the fidelity assessments of depolarising channels at individual hops. Distinctively, our routing policies prioritise paths based on the minimum required fidelity rather than maximum fidelity. We demonstrate that this approach enhances resource utilisation significantly, setting our findings apart from those of Liu et al. \cite{Liu24}.

\subsection{Slotted model}\label{ssec:slotted_model}
Furthermore, we assume a connection-oriented model, where the applications on quantum devices must reserve the use of the quantum network through the controller, specifying the peer end device with which they wish to establish end-to-end entanglement of qubits and a minimum fidelity, which depends on the application's characteristics.
It is commonly understood that the use of connections unlocks the potential for higher utilisation of the (scarce) resources in the quantum network, especially when coupled with link-layer protocols for managing the local entanglement between quantum repeaters sharing a link \cite{jli22}.

\begin{figure}[tb]
\centering
\includegraphics[width=\columnwidth]{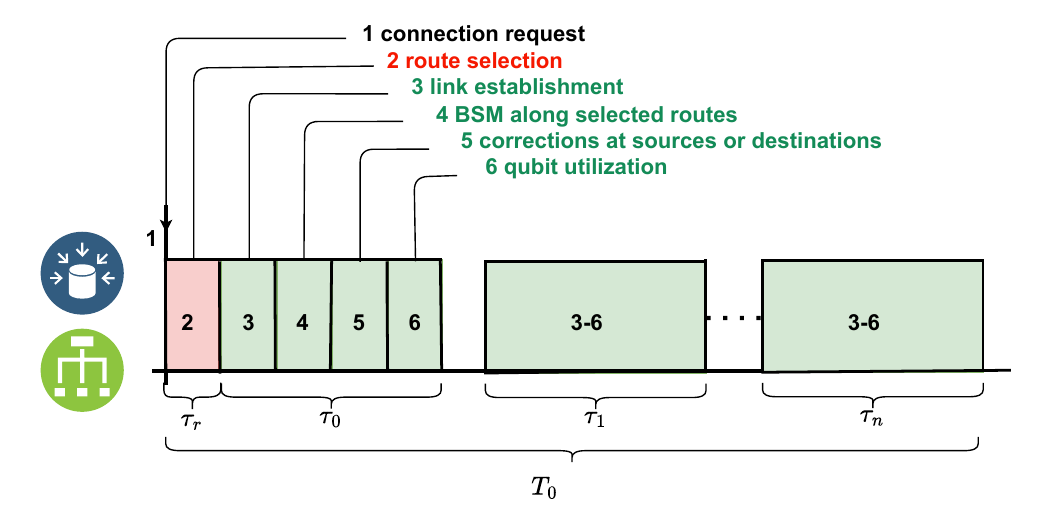}
    \vspace{-0.6cm}
    \caption{The timing diagram}
    \label{fig:time}
\end{figure}

In our system, we model the time evolution in \emph{periods}, each marked by a change in the set of active connections.
In other words, a new period begins ($T_0$ in Figure~\ref{fig:time}), when either a new connection requests to be admitted by the controller or a previously active connection is terminated by the end devices because the application has been completed.
Upon starting a new period, the controller is responsible for assigning a path to the active (and new) connections, using a route selection algorithm such as that in Algorithm~\ref{alg:routing}. This phase is represented with duration \( \tau_{r} \) in Figure~\ref{fig:time}. During this phase, all the quantum repeaters are also provided with a periodic schedule of instructions on how to perform entanglement swapping to materialise the routing decisions made by the controller, e.g., adopting the architecture and procedures illustrated in \cite{skrzypczyk21}.
After that, and until the end of the period, the routes remain statically assigned for the connections that have been accepted, while those rejected (or \emph{blocked}) will not be able to make use of the quantum network in the current period.

Typically, it is assumed that during the period the system will evolve as a sequence of evenly-spaced time slots, each including the phases from 3 to 6 in Figure~\ref{fig:time}: establish local entanglement in a link (3), perform Bell state measurements (4), transfer the classical results to one of the quantum devices so that it can perform $\sigma_z / \sigma_x$ corrections (5), and consume the qubits on the quantum devices by the distributed applications running on them (6).
The duration of the time slot must be less than the decoherence time of the qubits, which with today's state-of-the-art technology is very challenging.
Furthermore, we note that local entanglements and Bell state measurements are stochastic processes, as they may fail with a given probability that depends on the specific technology used.
During the operation of the system, these failures may happen in Phases 3 and 4 and affect the long-term throughput, i.e., the number of end-to-end entangled qubits that can actually be used by the source and destination devices in the unit of time.
Instead, in the remainder of the paper, we will focus only on Phase 2, i.e., the route selection process, hence we can consider the stochastic failures orthogonal to our work.

\subsection{The Routing Problem}\label{ssec:routing}

We define the \underline{Routing Problem} as follows: \textit{For a quantum network modelled as a graph $G(V, E)$ with $G$ classes of quantum repeaters, where each $v \in V$ is a quantum repeater with efficiency figure $\eta_g$ ($g \in G$) and $e \in E$ is a link between two quantum repeaters, a set of connections of cardinality $n_{sd}$ requesting to use the network to create end-to-end entanglement of qubits between a source and a destination with a fidelity threshold $F_{th}$, find a set of non-overlapping paths that minimises the blocking probability while satisfying the fidelity requirement}.

We address the problem under the general framework illustrated by Algorithm~\ref{alg:routing}, by proposing several specific policies. The framework takes as input the quantum network topology represented as a graph \( G(V, E) \), the end-to-end (E2E) fidelity estimations, and a set of source-destination pairs \( \{s_k, d_k\} \) (where \( n_{sd} \) represents the total number of such pairs). Additionally, a path selection policy \( \mathcal{A} \) is employed to compute the optimal paths for the connections.

The algorithm begins by initializing the set of assigned paths \( P_a \) to an empty set (line 1). For each request to establish entanglement between source-destination pairs \( \{s, d\} \) (line 2), the policy \( \mathcal{A} \) is used to compute a feasible path \( \pi \) between the source and destination nodes in the network graph \( G(V, E) \), considering the network's current state and fidelity constraints (line 3). If a path \( \pi \) is successfully found, it is added to the set \( P_a \) (line 4). If no such path can be established, the connection between the source-destination pair \( \{s, d\} \) is marked as \emph{blocked} for that period (line 5).

The process repeats for all \( n_{sd} \) pairs, and at the end of the algorithm, the final set of assigned paths \( P_a \) is returned as the output (line 7).
The policies are as follows:\newline
\paragraph{\underline{Shortest-path approach}} \mbox{}\\
This policy entails the straightforward use of Dijkstra's algorithm i.e., selects one of the shortest paths between source and destination from the set of those with fidelity above the threshold. When more such shortest paths exist, one is picked randomly with uniform distribution. Considering the generic path $p$ characterised by a path length $L_p$, an end-to-end fidelity $F_p$, and considering a fidelity threshold $F_{th}$, the path selection under the shortest-path approach selects the best path $P_a$ as follows:
\begin{equation}\label{eq:sp} 
P_a = \arg\min_{p} \{L_p \mid F_p \geq F_{th}\}.
\end{equation}
\begin{algorithm}
\caption{Route selection algorithm}
\label{alg:routing}

\textbf{Input}:
Network as a graph $G(V,E)$, E2E fidelity estimations, source-destination pairs $\{ s_k, d_k \}, k = 1..n_{sd}$, where $n_{sd}$ is the number of connections, and $\mathcal{A}$ is the path selection policy

\textbf{Output}:
Set of assigned paths $P_a$

\begin{algorithmic}[1]
\STATE $P_a$ = $\emptyset$
\FOR {each pair $\{ s, d \}$}
\STATE compute the path $\pi$ using policy $\mathcal{A}$ between $s$ and $d$ in $G(V,E)$
\STATE if $\pi$ is found, add it to $P_a$
\STATE otherwise the connection with pair $\{s,d\}$ is \emph{blocked} in this period
\ENDFOR
\STATE return $P_a$
\end{algorithmic}
\vspace{-0.3em}
\end{algorithm}
\vspace{2mm}
\paragraph{\underline{Knowledge-aware approach}} \mbox{}\\
This policy uses additional knowledge of quantum repeaters i.e., their efficiency figures. Therefore, it is a white-box approach solution to the quantum routing problem. Each node $v$ in the quantum network is assigned a cost $c(v)$ according to its efficiency figure. Then a vertex-weighted Dijkstra's algorithm is employed for path-selection. A preliminary study of such a policy was done in~\cite{vk23}. Path selected:

\begin{equation}\label{eq:ka} 
P_a = \arg\min_{p} \{C_p = \sum_{v \in P} c(v) \mid F_p \geq F_{th}\}.
\end{equation}
\vspace{2mm}
\paragraph{\underline{$k$-shortest path approach}} \mbox{}\\
This policy selects the path with minimum end-to-end fidelity (yet above fidelity threshold) out of first $k$ shortest-paths between source and destination list, where $k$ is a configuration parameter of the algorithm. Note that those paths may be of different lengths (not all equal to the shortest one). The intuition behind this policy is that selecting paths with a lower fidelity, corresponding to quantum repeaters in low-quality classes, in the early rounds of the loop in Algorithm~\ref{alg:routing} should leave more resources available to connections under evaluation later on.
Since no per-node/per-link information is needed, the $k$ shortest-path is a grey-box approach solution to the quantum routing problem. A more descriptive name for this algorithm could be ``min-k-shortest path", as it clarifies that the algorithm does not involve random selection among shortest paths or selection of multiple paths. However, for simplicity, we retain the term ``$k$-shortest path" throughout the paper. Path selected:
\begin{equation} \label{eq:ksp}
P_a = \arg\min_{p} \{F_p \mid F_p \geq F_{th}\}.
\end{equation}
\vspace{2mm}
\paragraph{\underline{$kx$-path selection approach}} \mbox{}\\
This policy is another grey-box approach solution to the quantum routing problem and is a restrictive version of the $k$-shortest path approach: when selecting the path, it only considers candidate routes that do not deviate too much from the shortest possible one, depending on the configuration parameter $x$, which can be seen as an allowance factor. When $x = 0$ the path must be no longer than the shortest available, when $x = 1$ it is allowed to be one hop longer than the shortest one, and so on. Let us define the length of the shortest path as follows:

\begin{equation} \label{eq:kxb}
L_b = \min_{p} \{L_p \mid F_p \geq F_{th}\}.
\end{equation}

Path selected:
\begin{equation} \label{eq:kxa}
P_a = \arg\min_{p} \{F_p \mid L_p \leq (L_b + x) \wedge F_p \geq F_{th}\}.
\end{equation}
If more than one path satisfies Equation~\ref{eq:kxa}, the algorithm selects a random one with a uniform distribution.



\section{Simulation Methodology}\label{sec:sim_and_method}
We developed a specialised Python-based simulator to compare the performance of the algorithms described in Section~\ref{ssec:routing}, when applied to a network as described in Section~\ref{ssec:architecture}. Specifically, we aim at characterising the fidelity achieved by a target set of source-destination pairs, depending on the different routing strategies, when applied to phase 2 of Figure~\ref{fig:time}. As will be clear from the experimental results, the order in which paths are computed for the different source-destination pairs in the set may determine the achieved fidelity for a specific source-destination pair. Therefore, we use Monte Carlo simulations to generate random orders in which we compute the best paths for a given set of source-destination pairs, and average the results to remove the dependence on the position in the order.


\begin{figure*}
\centering
    \begin{subfigure}{0.52\textwidth}
        \includegraphics[width=\linewidth, trim={3cm 3cm 3cm 3cm}, clip]{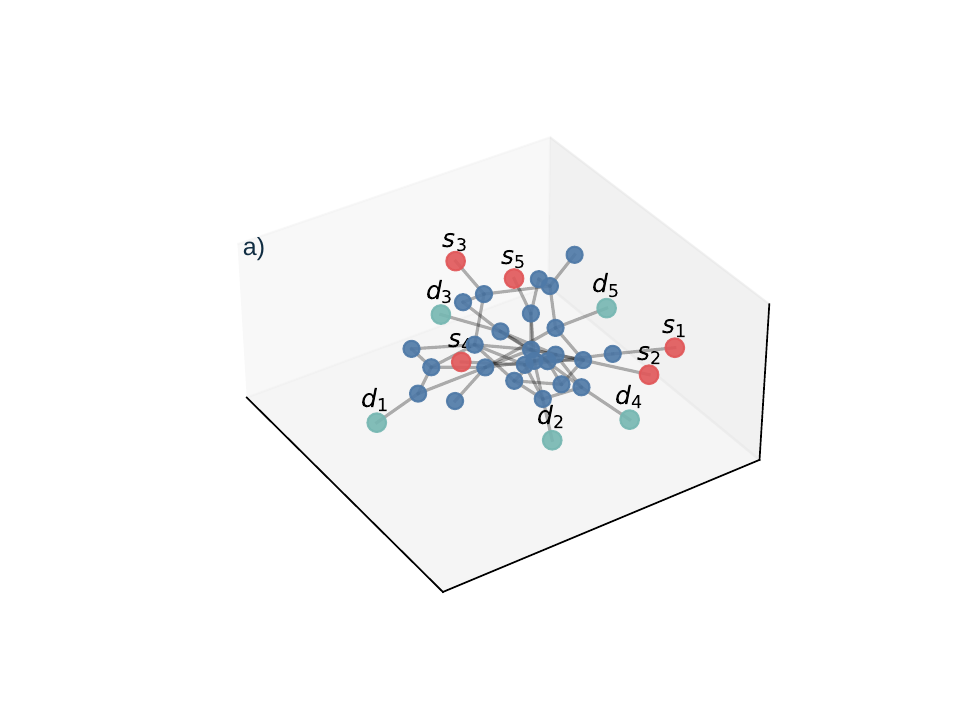}
        \caption{Random}
    \end{subfigure}
    \hspace{-1cm}
    \begin{subfigure}{0.52\textwidth}
        \includegraphics[width=\linewidth, trim={3cm 3cm 3cm 3cm}, clip]{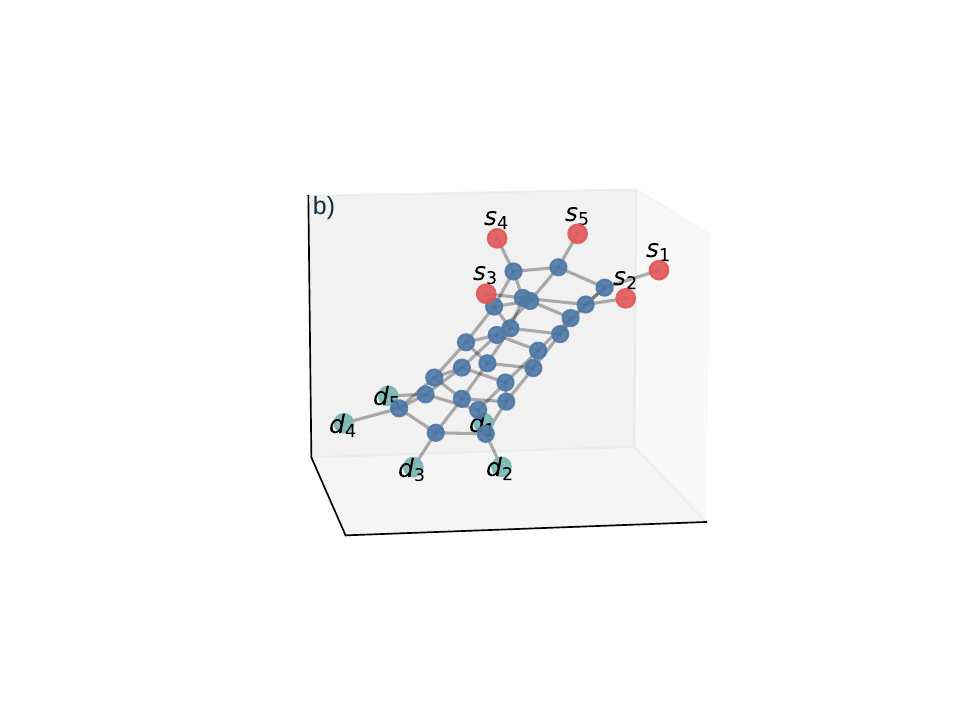}
        \caption{Regular}
    \end{subfigure}

    \caption{Topology of the quantum network for \( n_{sd} = 5 \) and 25 grid nodes. Source nodes \( s_n \) (in red), destination nodes \( d_n \) (in light blue), and grid nodes (in dark blue). In a regular network structure, the source-destination (SD) pairs are positioned farther apart, leading to a probability mass function (PMF) with less diversity in path lengths and a higher proportion of longer paths compared to a random network.
}
    \label{fig:topology}
\end{figure*}

\subsection{Topology and devices}\label{topology_and_devices}
With respect to the generic network shown in~Figure~\ref{fig:network_architecture}, in the following we focus on two specific topologies, referred to as \emph{regular} and \emph{random}, respectively as shown in Figure~\ref{fig:topology}. In the regular topology, the quantum repeaters are arranged in a $n \times n$ square grid with a connection between top and bottom transport nodes to avoid any edge effects. The source and destination nodes are located on the opposite edge of the transport network grid. In the random topology, we use the well-known Waxman model~\cite{wax88} to generate the specific network topology. Specifically, we use the Waxman model to generate the network connecting the quantum repeaters among them. Then, we select $n_{sd}$ repeaters using a uniform random distribution and attach them to the source nodes, and another $n_{sd}$ repeaters to attach the destination nodes.
When generating the network among repeaters according to the Waxman model, a link between any two nodes $u$ and $v$ is added with the following probability:

\begin{equation} \label{eq:wax}
P(u, v) = \beta e^{\left( \frac{-d(u, v)}{L \alpha} \right)}
\end{equation}
where, $d(u, v)$ represents the distance from node $u$ to node $v$, $L$ denotes the maximum distance between any two nodes, and $\alpha$ and $\beta$ are parameters that fall within the range $(0, 1)$. Higher values of $\beta$ lead to graphs with greater edge densities. Conversely, lower values of $\alpha$ tend to increase the proportion of shorter edges compared to longer ones. The Waxman model is one of the models more commonly used to generate quantum network topologies in simulations.

We initially consider two classes of quantum repeaters. Depending upon the efficiency of the quantum repeater, the node is categorised as high-quality (HQ), having efficiency $\eta_h$, or low-quality (LQ), with efficiency $\eta_l$. For what KA is concerned, we assign the costs so that $c(v_{\mathrm{HQ}}) \gg c(v_{\mathrm{LQ}})$.  In the first set of experiments, we vary the fraction of HQ and LQ repeaters. Specifically, a parameter $\xi$ is defined as the fraction of high-quality nodes among all quantum repeaters i.e., $\xi = 1 (\xi = 0)$ means that all quantum repeaters are HQ (LQ). Moreover, to extend the analysis with respect to this standpoint, in another set of experiments we allocate efficiency values over a continuous interval between the HQ and LQ values, according to the following considerations. If we selected efficiencies of repeaters according to a uniform distribution in this interval, as the end-to-end fidelity exponentially decays with respect to the efficiency figures (Equation~\ref{eq:fidelity}), the fidelity of end-to-end paths would be predominantly influenced by the efficiency figures closer to LQ. Therefore, to obtain a better balance, we use the following algorithm to assign efficiency values. For a given repeater $i$, we use two ancillary parameters ($a$ and $x_i$) and write the efficiency $\eta_i$ as follows:

\begin{equation}
\eta_i = \frac{\ln \left( x_i  \right)}{a},
\end{equation}
where $a$ is a constant that tunes the bias towards higher efficiency values ($a=1$ means unbiased) and $x_i$ is drawn from a uniform distribution in the following range:

\begin{equation}\label{eq:range}
U\left( e^{a \eta_{l}}, e^{a \eta_{h}} \right).
\end{equation}
%

Finally, a parameter $\theta$, path establishment order, is defined that represents the order (from 1 to the number of source-destination (SD) pairs considered in the experiments, $n_{sd}$)  of a specific SD pair in the sequence used to serve all pairs. 

\subsection{Performance figures}
\label{sub:performance_figures}

The performance of the proposed algorithms is evaluated based on the following set of indices.

First, we consider the \emph{Blocking Probability} (BP) across the entire set of considered SD pairs, computed as the fraction of SD pairs to which the routing algorithm assigns a path, over the total number of SD pairs requesting connectivity. A lower blocking probability indicates better network performance. In addition, for some specific analyses, we also consider the Blocking Probability per edge (BP/E) defined as the Blocking Probability divided by the number of edges in the network. As with blocking probability, a lower BP/E reflects improved network performance. Additionally, we analyse the probability mass function (PMF) of path length, which describes the probability distribution of the path lengths selected to connect SD pairs. PMF varies with the routing approach, the network parameter $\xi$, and the network topology.

A second index we use is the \emph{fidelity} achieved by SD pairs over the paths allocated by the routing algorithm, computed as in Equation~\ref{eq:fidelity}. We also consider the path establishment order ($\theta$), which represents the sequence in which SD pairs are served. In an optimised approach, the path order should minimally impact network performance. However, as shown in the results, this behaviour may not hold, particularly when the number of source-destination (SD) pairs increases, which can reduce network performance.

Finally, we characterise the \emph{fairness} achieved by the served SD pairs, as those served later may obtain ---if the routing algorithm is not cautious--- quite lower resources than those served first. To this end, we use Jain's fairness index~\cite{jain84} computed over the number of paths successfully allocated to each SD pair across all replicas of a given experimental configuration, i.e.:

\begin{equation} \label{eq:jain}
J(x_1, x_2, ..., x_n) = \frac{\left( \sum_{i = 1}^{n} x_i \right)^{2}}{n \sum_{i = 1} n x_{i}^{2}}
\end{equation}
where, $x_i$ is the count of path delivered to the $i_{th}$ (source, destination) pair.

\section{Results}\label{sec:results}

\begin{table}[h!]
\centering
\caption{Notation Table}
\label{tab:notation_table}
\begin{tabular}{cp{5cm}p{1.5cm}}
\toprule
Notation & Explanation & Default value \\
\midrule
$\xi$ & Fraction of high-quality nodes in the network &  \\
$\eta_h$ & Efficiency figure of a high-quality node & 0.999 \\
$\eta_l$ & Efficiency figure of a low-quality node & 0.8 \\
$F_{th}$ & Threshold fidelity &  0.53\\
$F_p$ & End-to-end fidelity of the path &  \\
$F$ & Fidelity of initial EPR pairs shared & 0.975 \\
$L_p$ & Path length i.e., number of nodes in the path &  \\
$\alpha$ & Waxman parameter that controls the proportion of shorter edges compared to longer &  0.85\\
$\beta$ & Waxman parameter that controls the edge density i.e., average degree & 0.275 \\
$k$ & Algorithm parameter for number of candidate shortest paths & 10 \\
$\theta$ & Path establishment order &  \\
\bottomrule
\end{tabular}
\end{table}
In this section, we evaluate the performance of the system presented in Section~\ref{sec:net_and_mot}, and specifically the performance of different routing approaches presented in Section~\ref{ssec:routing}. 
 Throughout this study, a quantum network consisting of 25 quantum repeaters is used irrespective of topology and number of source-destination pairs involved. A total of 100 random combinations of source and destination nodes are considered in addition to 100 random assignations of HQ and LQ nodes to the transport network. This leads to a total of $10,000 \times n_{sd}$ potential paths i.e., $50,000$ for $n_{sd} = 5$ for each $\xi$ value. The performance results for each $\xi$ are obtained by averaging over random assignation of HQ and LQ along with t-Student's 95\% confidence intervals. The results will only be shown for $0.5< \xi < 1$ as having quantum network with $\xi< 0.5$ only further decreases network performances since it would involve more lower quality nodes than higher quality nodes. The fidelity of the initially shared EPR pairs should be as high as possible, ideally close to 1, to tolerate the reduction in fidelity caused by noisy entanglement swapping procedures, thus ensuring that the entanglement shared between the source-destination pair remains above the fidelity threshold. However, achieving a fidelity of 1 is unrealistic; therefore, we assume the initial EPR pairs shared among neighbouring nodes have a fidelity of 0.975. We set the efficiency figures for high-quality and low-quality nodes at $\eta_h = 0.999$ and $\eta_l = 0.8$, respectively. It is important to note that the typical fidelity required by quantum applications should exceed 0.5. Consequently, we establish the threshold fidelity required for quantum applications at 0.53. However, applications running on end nodes may demand higher fidelity levels. Our choice of a lower fidelity threshold enables clearer differentiation of algorithmic performance and enhances the interpretability of the results. Notably, as observed in a prior study \cite{vk23}, algorithm performance may converge at higher fidelity thresholds. For readers interested in outcomes based on these higher thresholds, we refer them to our preliminary analysis where we explored this phenomenon. Additionally, we set $k = 10$ for both the $k$-shortest path and $kx$ path-selection approaches, providing a sufficient number of shortest paths from which to select. The results are categorised into six categories i.e., blocking probability, fairness, number of source-destination pairs, average degree, robustness, and range of efficiency figures. For each of these metrics, we evaluate the performance of all the path selection policies defined in Section~\ref{ssec:routing}: shortest-path (SP) and knowledge-aware (KA), as baselines, and $k$-shortest path (KSP) and $kx$-path selection (with $x=0$ as $kx_0$ and $x=1$ as $kx_1$), as the proposed novel approaches.
 \begin{figure}[tb]
    \includegraphics[width=\columnwidth]{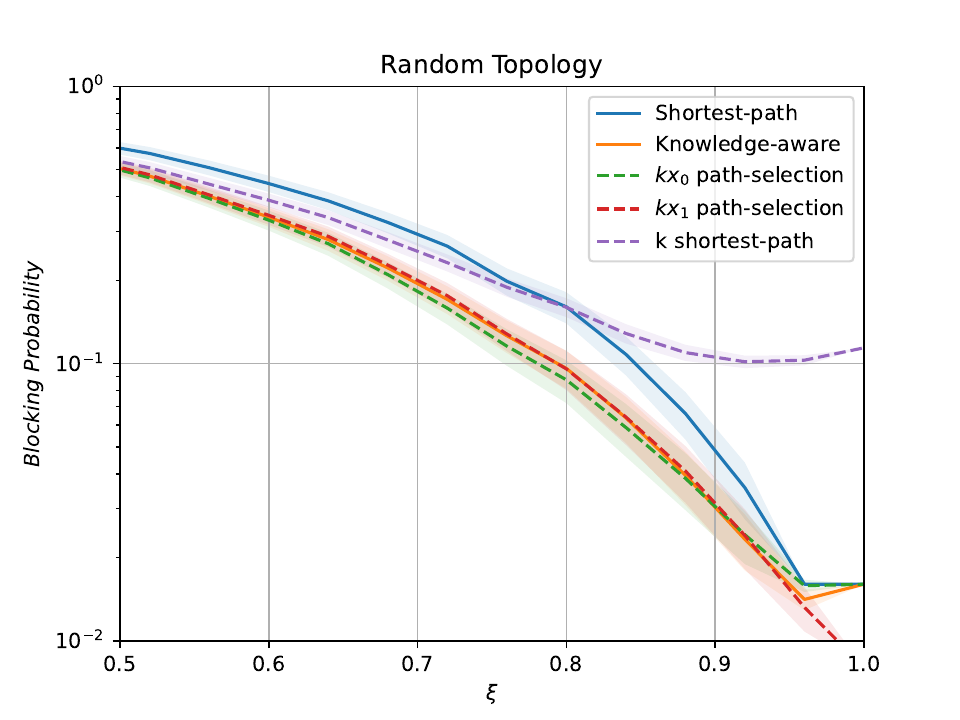}
    \vspace{-0.6cm}
    \caption{Blocking probability (BP) as a function of the fraction of high-quality nodes ($\xi$) in a randomized transport network. Parameters are set with a fidelity threshold ($F_{th}$) of 0.53 and a source-destination count ($n_{sd}$) of 5.}
    \label{fig:bp_wax_conf}
\end{figure}
\begin{figure}[tb]
    \includegraphics[width=\columnwidth]{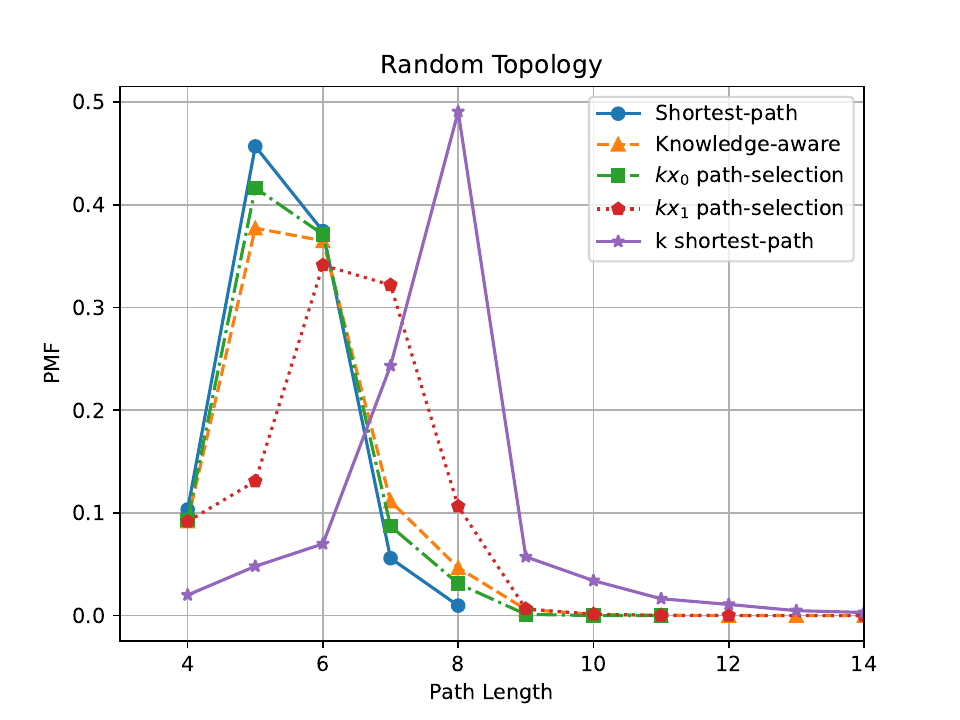}
    \vspace{-0.6cm}
    \caption{Probability mass function (PMF) of path length in a randomized transport network with a fidelity threshold ($F_{th}$) of 0.53 and a source-destination count ($n_{sd}$) of 5.}
    \label{fig:pl_wax}
\end{figure}
\begin{figure}[tb]
    \includegraphics[width=\columnwidth]{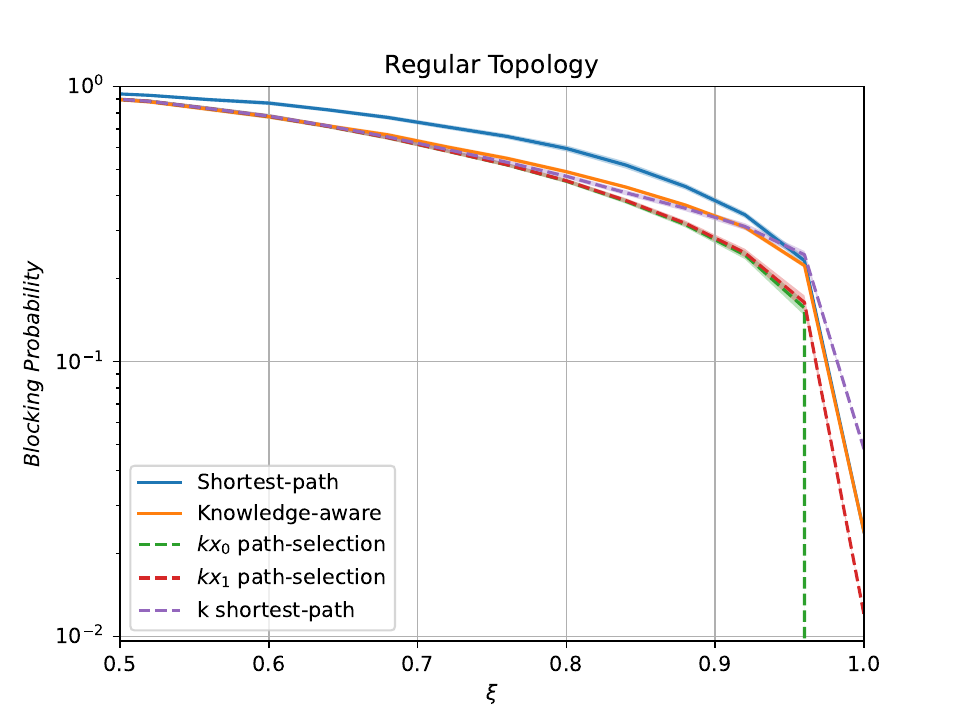}
    \vspace{-0.6cm}
    \caption{Blocking probability (BP) as a function of the fraction of high-quality nodes ($\xi$) in a regular transport network. Parameters are set with a fidelity threshold ($F_{th}$) of 0.53 and a source-destination count ($n_{sd}$) of 5.}
    \label{fig:bp_grid_conf}
\end{figure}
\subsection{Blocking Probability}\label{ssec:bp}
In this aspect, the methodology described in Section~\ref{sec:sim_and_method} is applied to both random and regular topology with $n_{sd}$ = 5. The parameters of the Waxman model are tuned to have a network of the same number of nodes and links as the regular topology, to compare the performance over the two networks on fair grounds.

\textit{\underline{Random Topology:}} Figure~\ref{fig:bp_wax_conf} displays the trend of blocking probability versus the fraction of HQ repeaters ($\xi$) for random topology and the corresponding PMF of the path lengths in Figure~\ref{fig:pl_wax}. The three evaluated approaches i.e., $kx_{0}$, $kx_{1}$, and KA outperform the baseline SP approach, represented in blue solid line. Although these three approaches demonstrate nearly identical blocking probabilities and PMF with SP approach, the $kx_{0}$ approach is the most optimised. While KA requires the knowledge of network elements' characteristics to reduce the blocking probability, $kx_{0}$ achieves these results without it as described in Section~\ref{sec:net_and_mot}. On the contrary, the KSP approach performed better than shortest-path until $\xi \leq 0.8$ and became worse for $\xi \geq 0.8$. This phenomenon can be attributed to the fact that while the KSP method selects paths based on optimising end-to-end fidelity, it tends to choose routes with greater path lengths, as illustrated in Figure~\ref{fig:pl_wax}. These longer paths consequently consume more valuable network resources for served paths, eventually resulting in higher blocking probabilities. This behaviour is also the reason why the blocking probability with KSP \emph{increases} for high values of $\xi$, i.e., when the quality of the quantum repeaters increases. This may seem counter-intuitive. Consider the extreme case of $\xi=1$, when all repeaters are of high quality. In this case, there would be a significant number of paths (of \emph{different} lengths) which are considered by KSP. KSP will pick the one with minimum fidelity (above the threshold), which therefore will be one of the longest ones among the candidate set. This will consume more qubits (with respect to picking a shorter one), eventually leading to higher blocking probabilities.

\begin{figure}[tb]
    \includegraphics[width=\columnwidth]{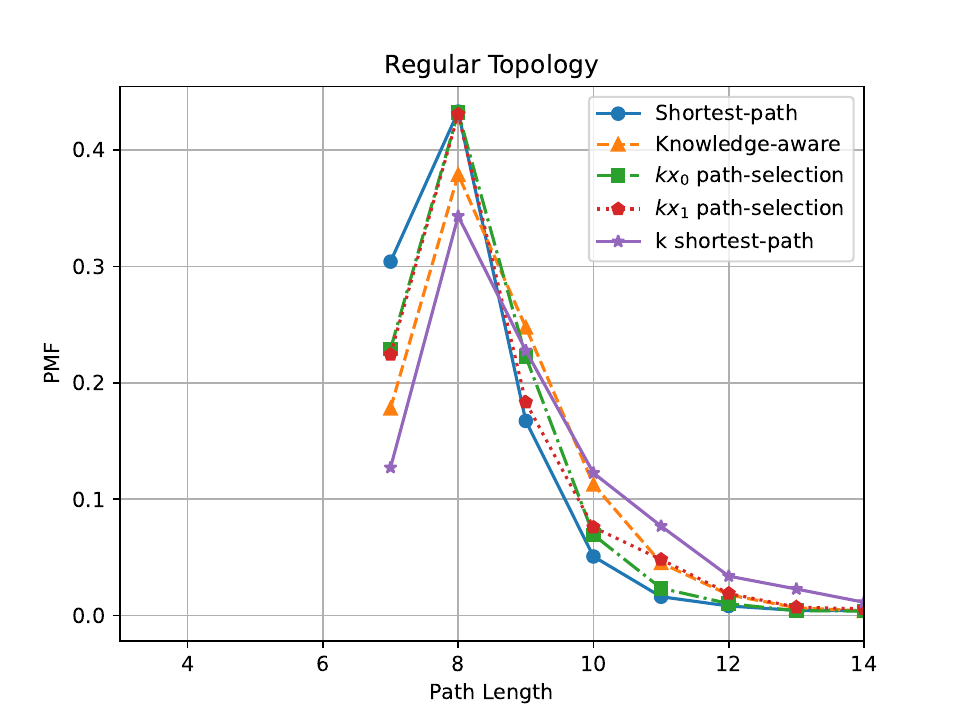}
    \vspace{-0.6cm}
    \caption{Probability mass function (PMF) of path length in a regular transport network with a fidelity threshold ($F_{th}$) of 0.53 and a source-destination count ($n_{sd}$) of 5.}
    \label{fig:pl_grid}
\end{figure}
\begin{figure}[tb]
    \includegraphics[width=\columnwidth]{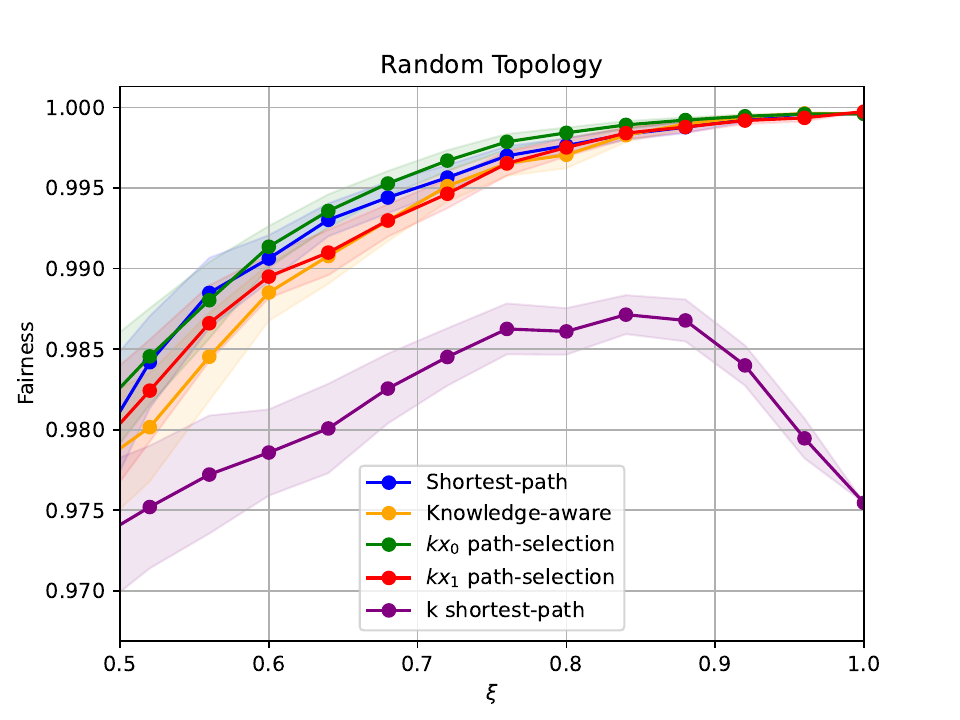}
    \vspace{-0.6cm}
    \caption{Fairness as a function of the fraction of high-quality nodes ($\xi$) in a randomized transport network, with a fidelity threshold ($F_{th}$) of 0.53 and a source-destination count ($n_{sd}$) of 5.}
    \label{fig:fair_wax_conf}
\end{figure}
\begin{figure*}[tb]
    \includegraphics[width=\textwidth]{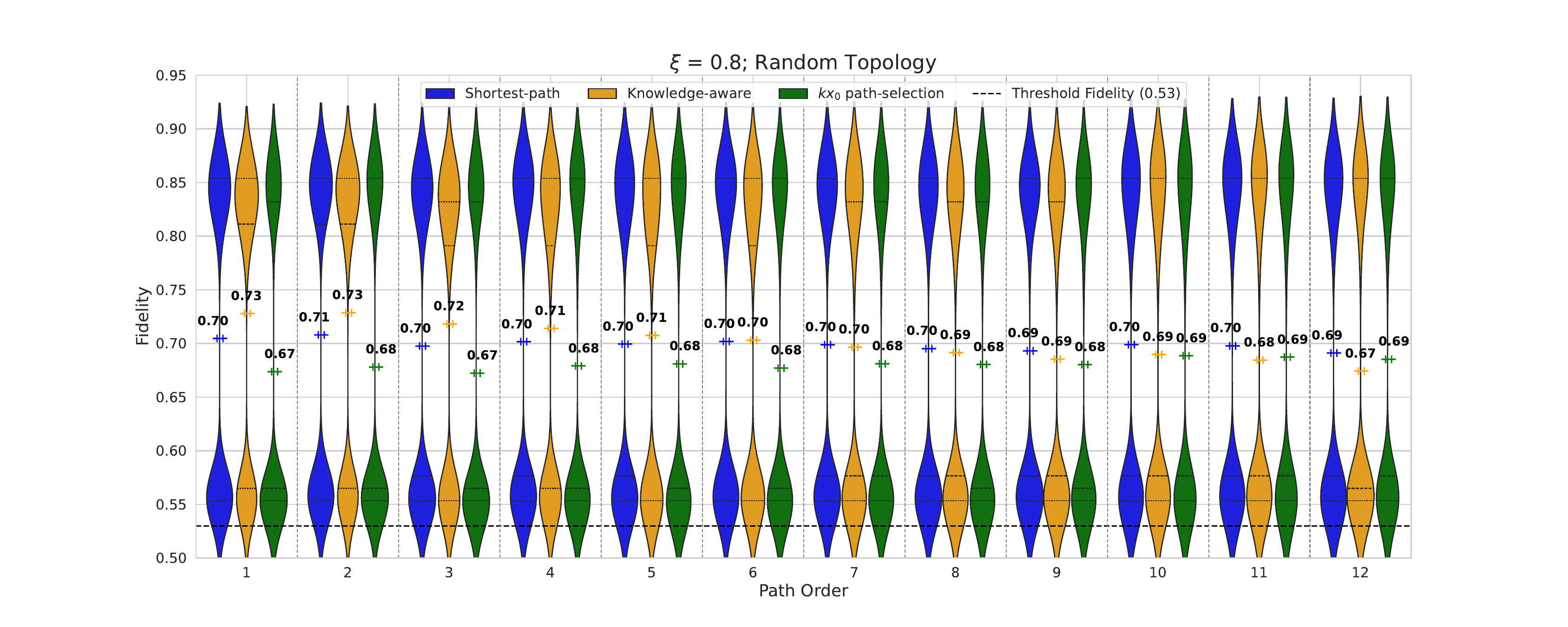}
    \vspace{-0.6cm}
    \caption{Fidelity as a function of path-order ($\theta$) in a randomized transport network, with a fidelity threshold ($F_{th}$) of 0.53 and a source-destination count ($n_{sd}$) of 12.}
    \label{fig:violin_wax_nsd12}
\end{figure*}
\textit{\underline{Regular Topology:}}
Figure~\ref{fig:bp_grid_conf} illustrates the trend in blocking probability as a function of $\xi$ for a regular topology, alongside the corresponding probability mass function (PMF) of path lengths depicted in Figure~\ref{fig:pl_grid}. $kx_{0}$, $kx_{1}$, KA, and KSP consistently outperform SP, except with $\xi=1$ when SP and KA become identical due to the cost of all nodes being the same. The network's regular structure predominantly results in a path length of 8, which emerges as the most frequent outcome, although variations are observed depending on the strategic approach adopted by each method. Consistent with performances observed in random topology, the $kx_{0}$ strategy proves to be superior to KA and KSP, especially with higher values of $\xi$. We note that KSP in regular topology does not suffer from significant performance degradation as with a random topology: this is because, due to the network structure in regular topologies, KSP has less freedom to explore paths that are significantly longer than the shortest one (see Figure~\ref{fig:pl_wax} vs.\ Figure~\ref{fig:pl_grid}).
\begin{table*}[h!]
\centering
\caption{Evaluation of Routing Performance}
\label{tab:routing_evaluation}
\begin{adjustbox}{width=1\textwidth}
\begin{tabular}{lccccc}
\toprule
\multirow{2}{*}{Property} & \multirow{2}{*}{Notation} & \multirow{2}{*}{Explanation} & \multicolumn{2}{c}{Performance Comparison} \\
\cmidrule{4-5}
& & & Random Topology & Regular Topology \\
\midrule
Blocking Probability & BP & Probability of a path being blocked & $kx_0$ $\sim$ $kx_1$ $\sim$ KA $>$ SP $>$ KSP & $kx_0$ $\sim$ $kx_1$ $>$ KSP $\sim$ KA $>$ SP \\
Fairness & J & Jain's index for multiple source-destination serving & $kx_0$ $\sim$ SP $>$ $kx_1$ $\sim$ KA $>$ KSP & SP $>$ $kx_0$ $\sim$ $kx_1$ $>$ KSP $>>$ KA  \\
Number of SD pairs & $n_{sd}$ & Number of source-destination pairs involved & $kx_0$ $>$ KA $\geq$ SP $>$ $kx_1$ $>$ KSP & N/A \\
\bottomrule
\end{tabular}
\end{adjustbox}
\end{table*}

Overall, the network topology exerts a significant influence on the performance outcomes of each approach. Blocking probability is notably higher in the regular topology compared to the random topology, attributed to the shifted path length distribution in the regular topology (see Figure~\ref{fig:pl_wax} \& \ref{fig:pl_grid}). This effect is evident in the comparative slopes of the blocking probability curves, with the regular topology in Figure~\ref{fig:bp_grid_conf} displaying a gentler slope than the random topology in Figure~\ref{fig:bp_wax_conf}. The network's regular structure limits the available route options because it offers little variation in served path lengths. This results in a relatively smaller improvement in performance as compared to the case in random topology.

\begin{figure}[tb]
    \includegraphics[width=\columnwidth]{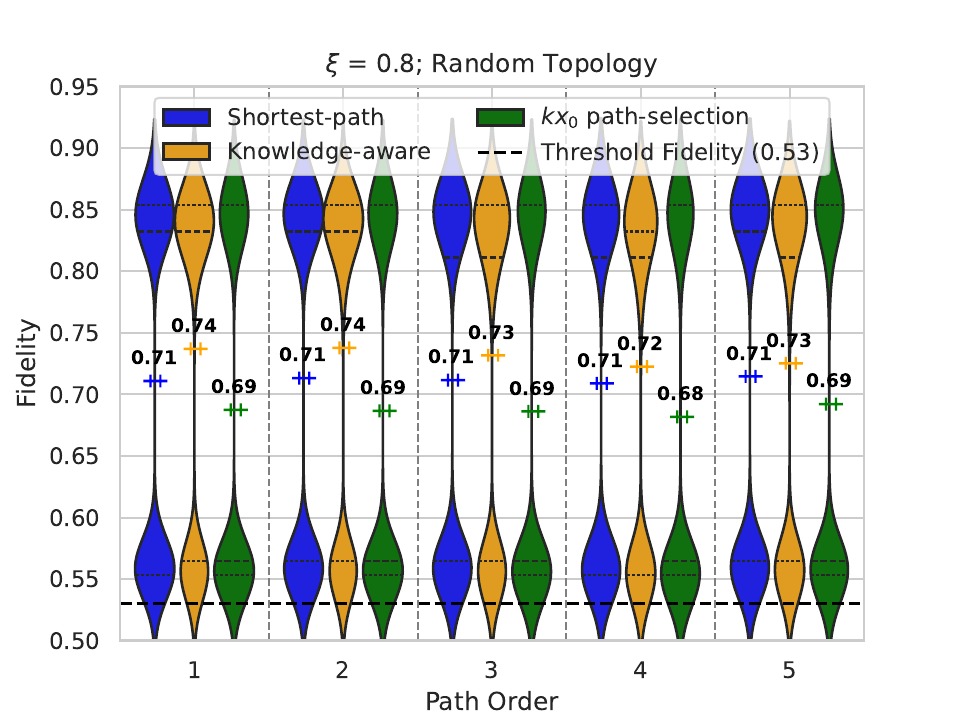}
    \vspace{-0.6cm}
    \caption{Fidelity as a function of path-order ($\theta$) in a randomized transport network, with a fidelity threshold ($F_{th}$) of 0.53 and a source-destination count ($n_{sd}$) of 5.}
    \label{fig:violin_wax}
\end{figure}
\begin{figure}[tb]
    \includegraphics[width=\columnwidth]{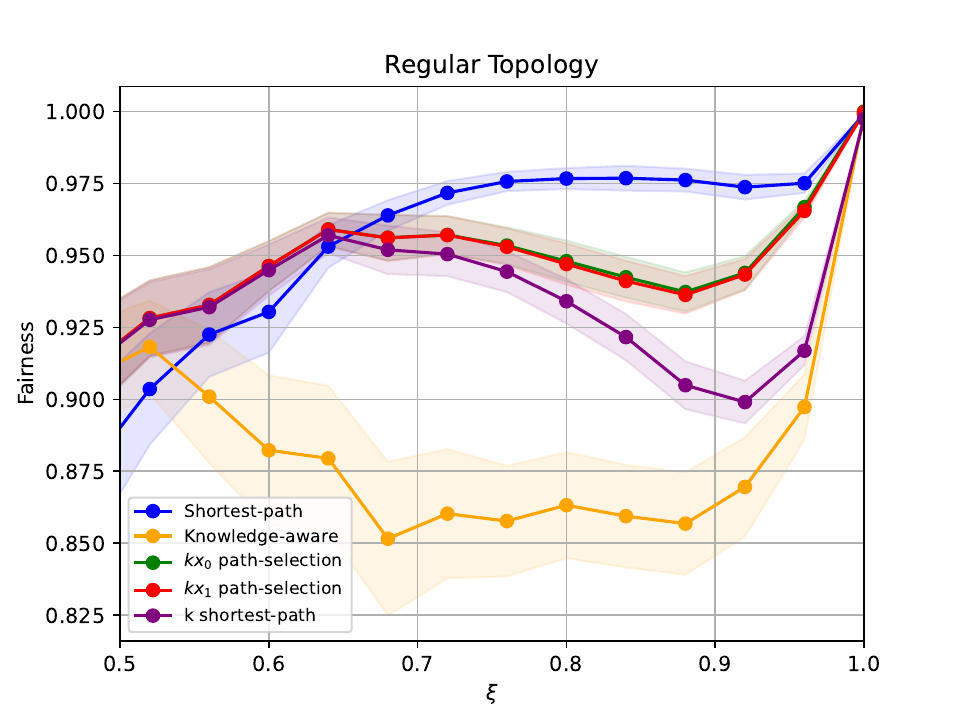}
    \vspace{-0.6cm}
    \caption{Fairness as a function of the fraction of high-quality nodes ($\xi$) in a regular transport network, with a fidelity threshold ($F_{th}$) of 0.53 and a source-destination count ($n_{sd}$) of 5.}
    \label{fig:fair_grid_conf}
\end{figure}
\begin{figure}[tb]
    \includegraphics[width=\columnwidth]{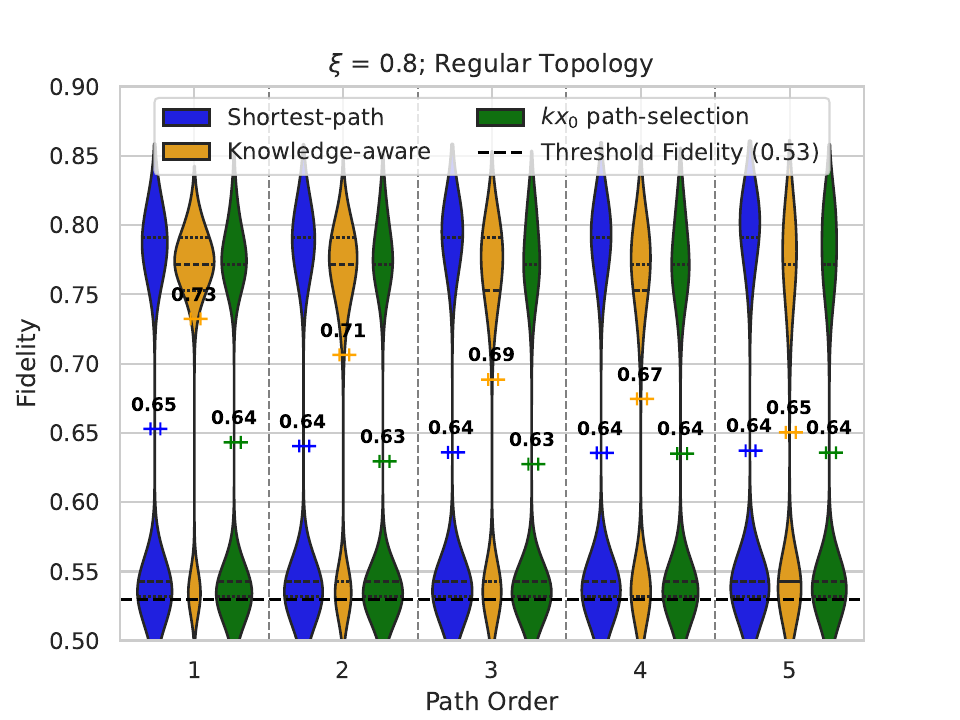}
    \vspace{-0.6cm}
    \caption{Fidelity as a function of path-order ($\theta$) in a regular transport network, with a fidelity threshold ($F_{th}$) of 0.53 and a source-destination count ($n_{sd}$) of 5.}
    \label{fig:violin_grid}
\end{figure}
\subsection{Fairness}\label{ssec:fairness}

\textit{\underline{Random Topology:}}
Figure~\ref{fig:fair_wax_conf} presents the fairness among network users, quantified using Jain's index, in terms of the number of paths successfully delivered above the fidelity threshold of 0.53 for the random topology.

The majority of the approaches evaluated exhibit comparable levels of fairness and increases their fairness as $\xi$ increases, with the notable exception of the KSP approach, which demonstrates significantly lower fairness and decreases its fairness as $\xi$ increases. This means that, in general, these policies are able to exploit higher quality of repeaters in a fair manner. The reason for the drop of fairness experienced by KSP for high $\xi$ values is related to the corresponding increase of the blocking probabilities observed in Figure~\ref{fig:bp_wax_conf} i.e., KSP using unnecessarily long paths as shown in Figure~\ref{fig:pl_wax} also for SD pairs that are topologically closer. In addition, those that are farther away (topologically) have thus a higher chance of being blocked, and this results in lower fairness values. Similarly, it is interesting to consider this result together with the performance in terms of blocking probability (Figure~\ref{fig:bp_wax_conf}) even for slightly lower values of $\xi$ i.e., $\xi < 0.8$. Although KSP has better BP in that range than SP, it does serve the non-blocked SD pairs fairly. The $kx_{0}$ approach not only maintains equitable treatment among users but also performs the best (among the considered approaches) in terms of BP (see Figure~\ref{fig:bp_wax_conf}). 

To better illustrate user fairness from another perspective, the fidelity distribution for each path order ($\theta$)—representing the distribution of served fidelity for requests in sequence (i.e., 1 to $n_{sd}$)—is presented. Figure~\ref{fig:violin_wax_nsd12} ($n_{sd} = 12$) and Figure~\ref{fig:violin_wax} ($n_{sd} = 5$) illustrate this fidelity distribution across path orders for various policies at $\xi = 0.8$, with the average fidelity served for each path order also highlighted. For presentation purposes we only plot $kx_{0}$ path-selection from the `grey box' approaches leaving behind worse performing KSP and $kx_{1}$ path-selection as discussed above. It is to be noted that $\xi = 0.8$ is chosen for analysis as the corresponding blocking probabilities are lower enough to be considered for realistic network settings. Since the policies are designed to deliver just over the threshold fidelity of 0.53, achieving higher fidelity than the fidelity threshold would be an over-utilisation of resources and can be considered as not fully optimised approach. The optimisation capability of $kx_{0}$ path-selection approach is visible from Figure~\ref{fig:violin_wax_nsd12} where it delivers a lower average fidelity (yet still over the fidelity threshold) for each path order with almost equal value as compared to other approaches. This shows that the approach uses low-quality nodes if the fidelity threshold allows for it which keeps the valuable resources to deliver the path for the upcoming path orders. The same is true for SP approach but the average fidelity values are more than the $kx_{0}$ path-selection approach which means that SP approach is not fully optimised. Contrary to this, KA has the worst case where it has unnecessarily higher average fidelity than the fidelity threshold for initial path orders and the average fidelity values gradually drops with path order. The same is true for $n_{sd} = 5$ as well in Figure~\ref{fig:violin_wax} but the effect is not clearly visible due to smaller number of source-destination pairs involved.

\textit{\underline{Regular Topology:}}
Figure~\ref{fig:fair_grid_conf} illustrates the fairness within the regular topology, while Figure~\ref{fig:violin_grid} shows the corresponding fidelity distribution for each path order ($\theta$) in the regular topology at $\xi = 0.8$. The average fidelity served for each path order is highlighted, similar to the random topology case. While all approaches exhibit comparable levels of fairness at $\xi = 0.5$, the KA method demonstrates the least fairness as $\xi$ increases. This trend is attributable to the preferential consumption of high-quality nodes in the initial delivery paths, a phenomenon also evident from Figure~\ref{fig:violin_grid}, where the average fidelity decreases almost linearly with increasing path order, represented in orange. Contrary to this, it is to be noted that KA did not exhibit worst fairness in random topology as seen in Figure~\ref{fig:fair_wax_conf} due to relatively shorter path length as seen in Figure~\ref{fig:pl_wax} than regular topology in Figure~\ref{fig:pl_grid}. The relatively longer path lengths, which arise from the inherent structure of the regular topology, enhance the effect of preferential consumption of high-quality nodes in regular topology.
Although the blocking probability of the KA policy is closer to other policies as in Figure~\ref{fig:bp_grid_conf}, it does not serve the paths fairly as in Figure~\ref{fig:fair_grid_conf}. 
\begin{figure*}[tb]
    \centering
    \begin{subfigure}{0.35\textwidth}
        \includegraphics[width=\linewidth]{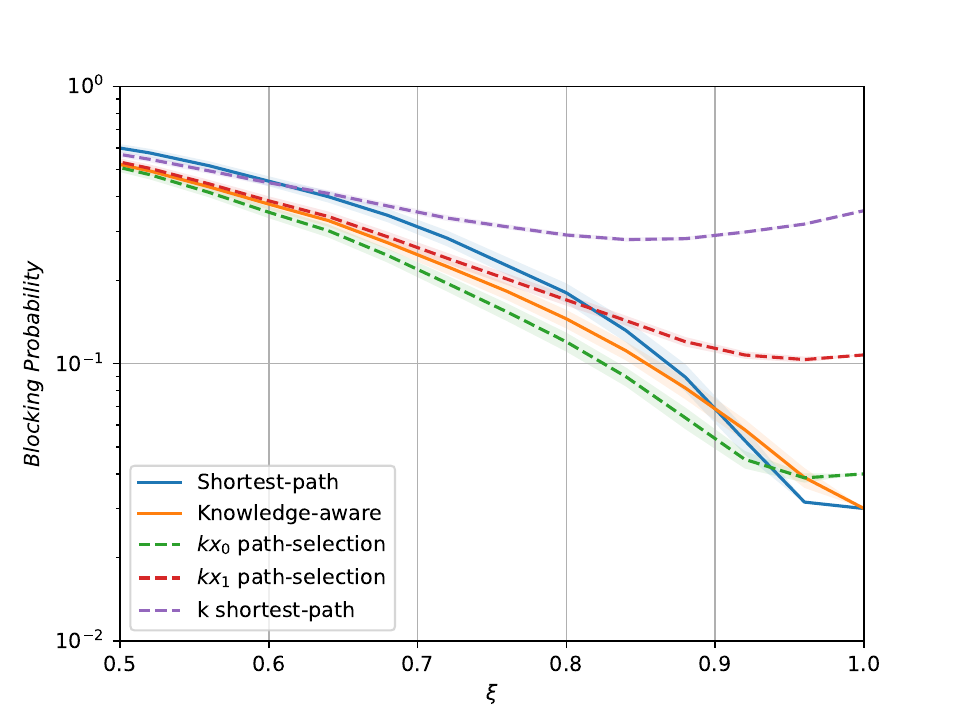}
        \caption{$n_{sd} = 8$}
        \label{fig:nsd8}
    \end{subfigure}
    \hspace{-0.7cm}
    \begin{subfigure}{0.35\textwidth}
        \includegraphics[width=\linewidth]{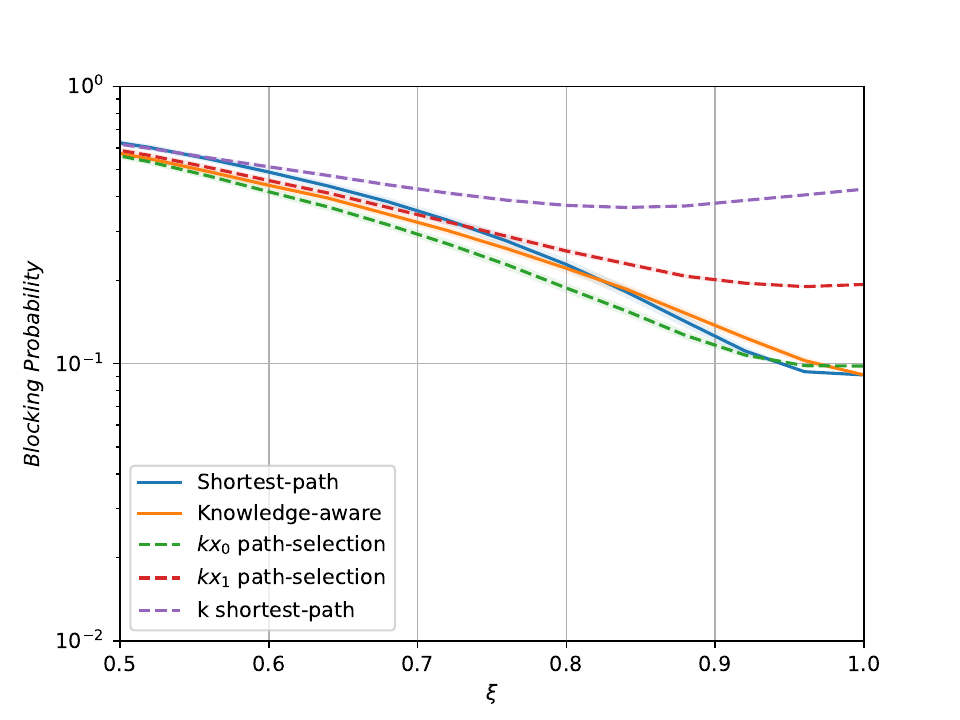}
        \caption{$n_{sd} = 10$}
        \label{fig:nsd10}
    \end{subfigure}
    \hspace{-0.7cm}
    \begin{subfigure}{0.35\textwidth}
        \includegraphics[width=\linewidth]{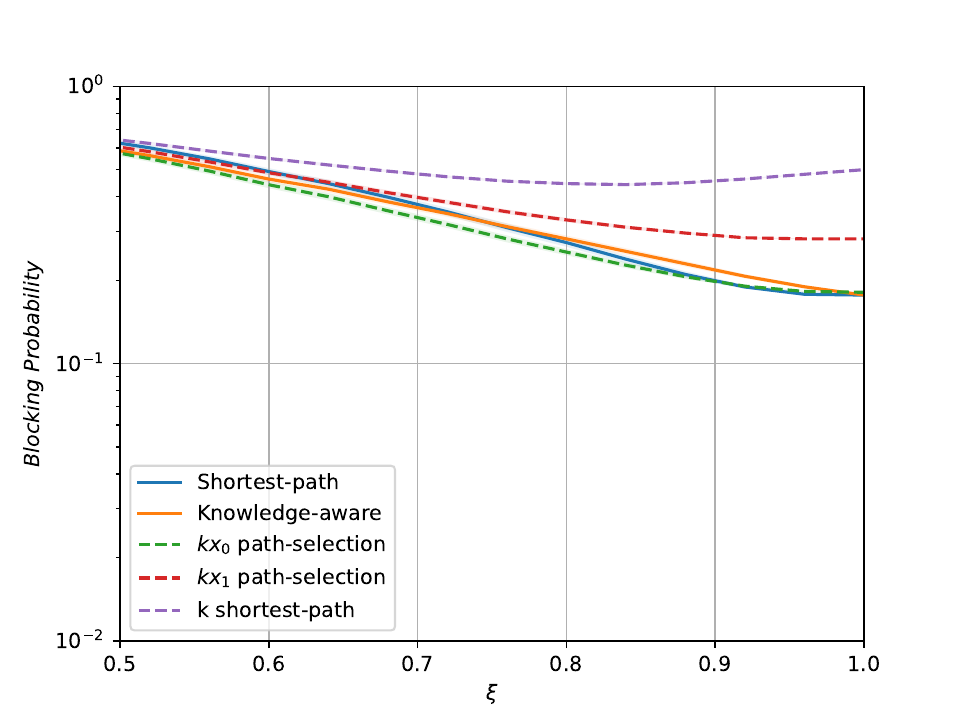}
        \caption{$n_{sd} = 12$}
        \label{fig:nsd12}
    \end{subfigure}

    \caption{Blocking probability (BP) as a function of the fraction of high-quality nodes ($\xi$) in a randomized transport network with a fidelity threshold ($F_{th}$) of 0.53, shown for source-destination counts ($n_{sd}$) of: a) 8, b) 10, and c) 12.}
    \label{fig:load}
\end{figure*}
An interesting observation in Figure~\ref{fig:fair_grid_conf} is the relation between blocking probability and fairness of the approaches. For mid to slightly higher value of $\xi$, approaches excluding KA and SP sacrifice a small amount of fairness as seen in Figure~\ref{fig:fair_grid_conf} to have a significant boost in the blocking probability performance as seen in Figure~\ref{fig:bp_grid_conf}. It is to be noted that this behaviour is only limited to regular topology and arises from the lack of availability of different path lengths as observed in Figure~\ref{fig:pl_grid} where only one peak is observed at path length of 8 while in case of random topology in Figure~\ref{fig:pl_wax}, path length of 5 \& 6 are observed almost equally. Ultimately, the $kx_{0}$ approach achieves commendable fairness in regular topology, surpassed only by the SP method which was observed to have significant less performance in terms of blocking probability.

\subsection{Number of source-destination pairs}\label{ssec:nsd}
We now analyse the impact of varying the number of source-destination pairs. For this analysis, we exclusively consider a random topology with a fixed number of transport nodes and links, and we vary the number of SD pairs, considering $n_{\text{sd}} \in \{5, 8, 10, 12\}$.


Figure~\ref{fig:bp_wax_conf} ($n_{sd} = 5$) and Figure~\ref{fig:load} ($n_{sd} \in \{8, 10, 12\}$) collectively demonstrates the network's performance under varying number of source-destination pairs for the random topology, displaying the blocking probability as a function of $\xi$. With increasing $n_{sd}$, the blocking probability escalates due to the limited availability of shared entangled pairs at quantum repeaters. Among the methods assessed, the $kx_{0}$ approach consistently outperforms all others across the different number of source-destination pairs. Notably, the $kx_{1}$ and KSP approaches, which perform well under lower number of source-destination pair, fail to maintain superiority over the SP approach as the $n_{sd}$ increases, exhibiting a decline in performance. In the case of KSP, the BP begins to increase with \( \xi \) when \( \xi \) approaches values close to 1 (i.e., \( \xi \in [0.9, 1] \)) for higher numbers of source-destination pairs (\( n_{sd} \in \{ 8, 10, 12\} \)). This counter-intuitive phenomenon arises from the KSP approach of selecting longer paths, which consumes valuable resources. This behaviour explained in Section~\ref{ssec:bp}, is further intensified by the increased load resulting from a higher number of SD pairs operating within the network. This shows that the algorithm $kx_{i}$ is very sensitive to allowing paths even longer only 1 hop more than the shortest possible, and this is more and more visible as the $n_{sd}$ increases. At $n_{sd} = 12$, the performance of SP, $kx_{0}$, and KA appears to converge, highlighting a uniform response to extreme network demands.

Table~\ref{tab:routing_evaluation} consolidates all results obtained thus far, providing a comprehensive summary of the performance comparison across the evaluated approaches, organized by the transport network topologies. The table includes the performance metrics discussed in Sections~\ref{ssec:bp}, \ref{ssec:fairness}, and \ref{ssec:nsd}, specifically blocking probability, fairness, and source-destination count.
\begin{figure}[tb]
\includegraphics[width=\columnwidth]{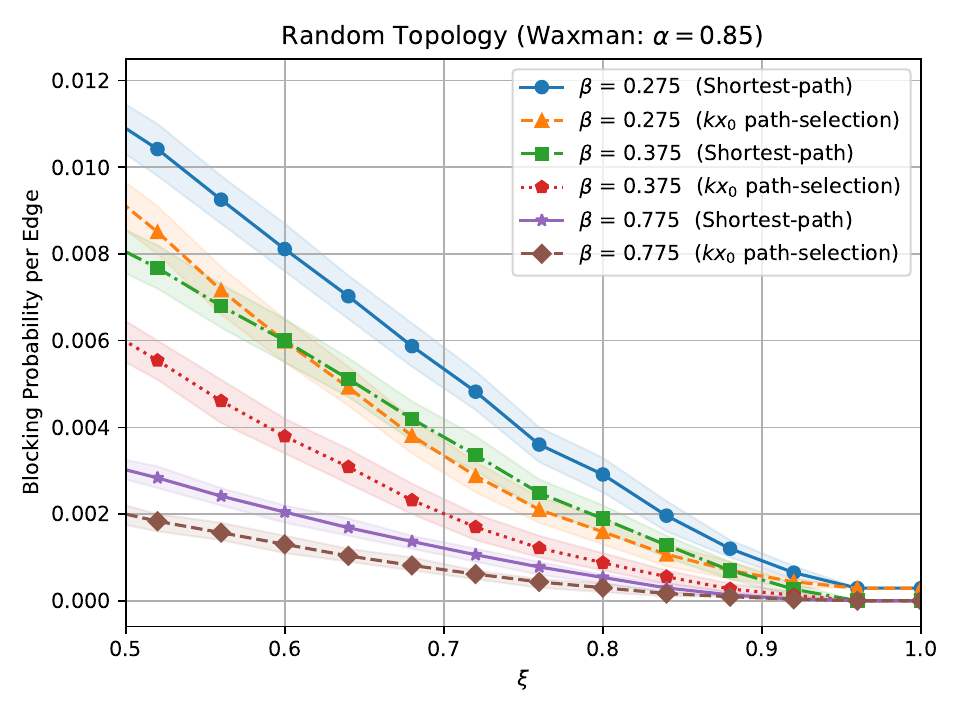}
    \vspace{-0.6cm}
    \caption{Blocking probability per edge (BP/E) as a function of the fraction of high-quality nodes ($\xi$) in a randomized transport network, with a fidelity threshold ($F_{th}$) of 0.53 and a source-destination count ($n_{sd}$) of 5, for varying Waxman parameter values ($\beta$).}
    \label{fig:bpe_wax_conf}
\end{figure}
\begin{figure}[tb]
    \includegraphics[width=\columnwidth]{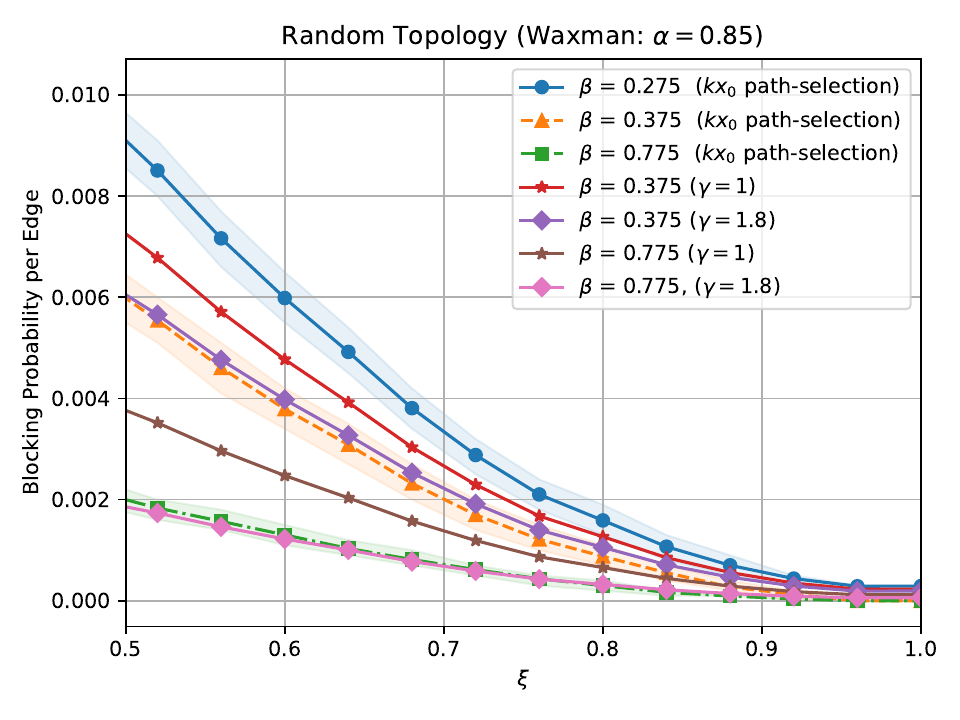}
    \vspace{-0.6cm}
    \caption{Blocking probability per edge (BP/E) as a function of the fraction of high-quality nodes ($\xi$) in a randomized transport network, with a fidelity threshold ($F_{th}$) of 0.53 and a source-destination count ($n_{sd}$) of 5, to model the performance of the $kx_0$ path-selection algorithm.}
    \label{fig:bpe_ref_wax_conf}
\end{figure}

\subsection{Average Degree}
So far, we have considered a random topology equivalent (in terms of number of nodes and links) to the reference regular topology. In this section, we investigate the effect of increasing the number of links (or average degree) in the network. To this end, we exploit the flexibility of the Waxman model in capturing the network density via the $\beta$ parameter. For the purposes of clarity in our presentation, we focus on evaluating the performance of the most effective routing approach identified in our study ---the $kx_0$ path selection method--- against the traditional shortest-path method for reference. We consider, as before, $n_{sd} = 5$ only.

Figure~\ref{fig:bpe_wax_conf} delineates the relationship between the BP/E and the parameter $\xi$ with a random topology. The figure clearly shows that as the network expands in complexity with an increased number of links, the $kx_0$ path selection method consistently outperforms the shortest path approach. Although not depicted in the figure, the $kx_0$ method also surpasses the other routing strategies under comparison.
We consider the BP/E instead of the BP because this
helps us better grasp how each new link addition individually affects the network. Specifically, we capture this effect by modelling the performance differentials between networks with varying average degrees ($\beta$) as eq~\ref{eq:bpe1}.

\begin{equation}\label{eq:bpe1}
    bpe_{\beta} = bpe_{\beta_{0}} \left( \frac{n_{e}^{\beta_{0}}}{n_{e}^{\beta}} \right)^{\gamma}
\end{equation}
where, $bpe_\beta$ is the BP/E for a random quantum network with Waxman parameters ($\alpha, \beta$) having $n_{e}^{\beta}$ number of links; $bpe_{\beta_{0}}$ is the BP/E for a random quantum network with Waxman parameters ($\alpha, \beta_{0}$) having $n_{e}^{\beta_{0}}$ number of links; $\gamma$ is a parameter that is selected to model the performance $bpe_\beta$ in terms of $bpe_{\beta_{0}}$.

As seen in Figure~\ref{fig:bpe_ref_wax_conf}, the value of $\gamma = 1.8$ can accurately predict the performance improvement of $kx_0$ path-selection approach in denser networks with $\beta = 0.375$ and $\beta = 0.775$ compared to the baseline case of $\beta = 0.275$, which entails a scalability that is slightly less than quadratic.
The results obtained in our simulated conditions show that, as network scales, the performance of $kx_{0}$ algorithm is given by: \newline
\begin{equation}\label{eq:bpe2}
    bpe_{\beta} = bpe_{\beta_{0}} \left( \frac{n_{e}^{\beta_{0}}}{n_{e}^{\beta}} \right)^{1.8}
\end{equation}

\begin{figure*}
\centering
    \begin{subfigure}{0.52\textwidth}
        \includegraphics[width=\linewidth]{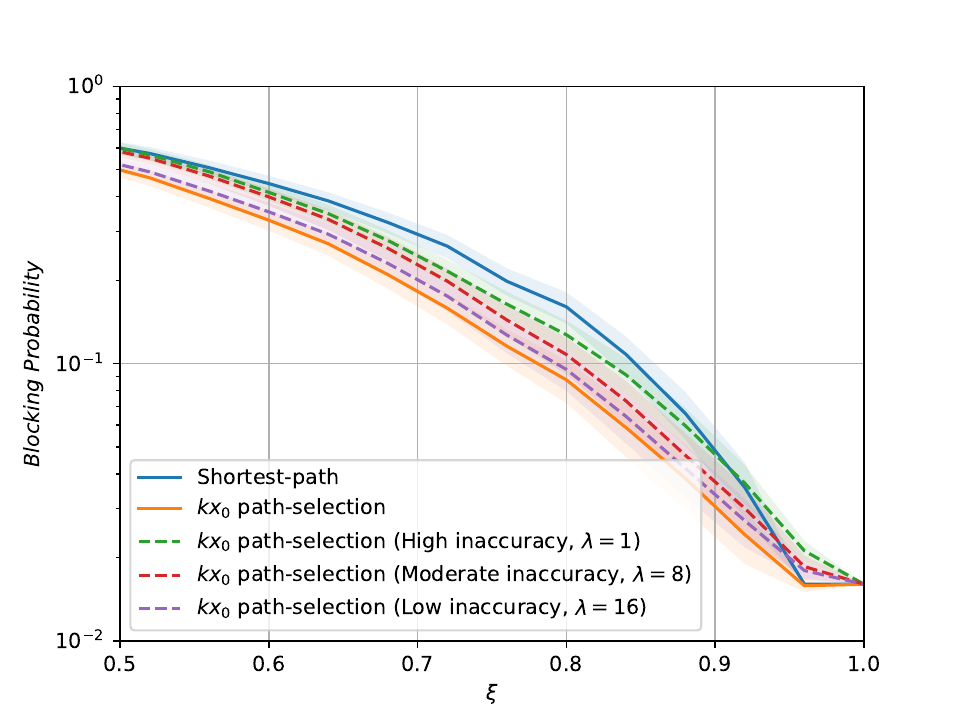}
        \caption{$n_{sd} = 5$}
    \end{subfigure}
    \hspace{-1cm}
    \begin{subfigure}{0.52\textwidth}
        \includegraphics[width=\linewidth]{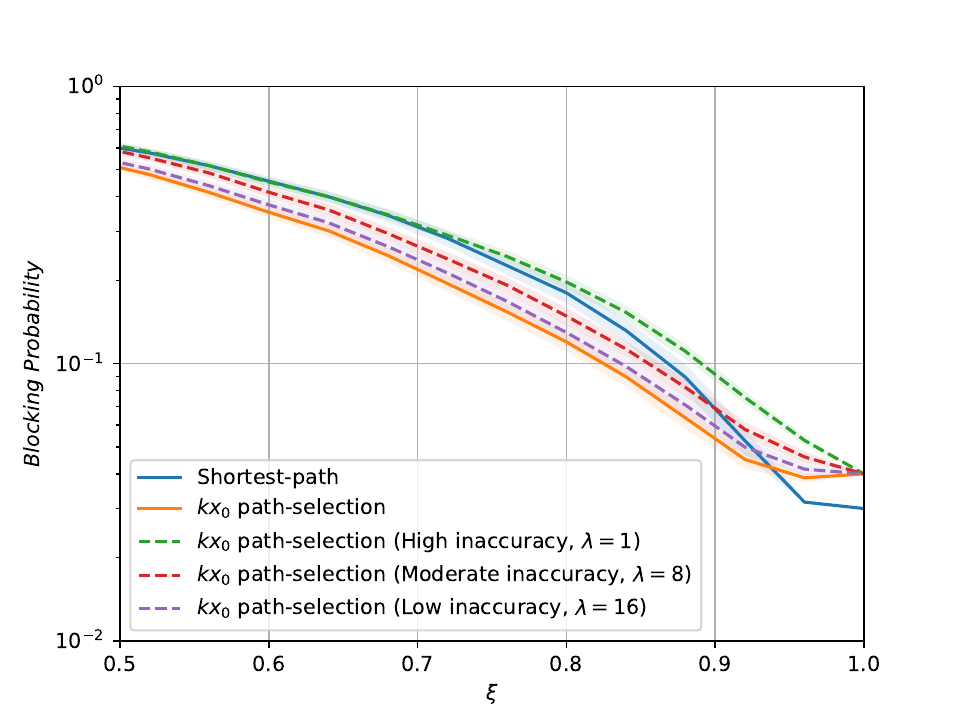}
        \caption{$n_{sd} = 8$}
    \end{subfigure}

    \caption{Robustness test: Blocking probability (BP) as a function of the fraction of high-quality nodes ($\xi$) in a randomized transport network, with a fidelity threshold ($F_{th}$) of 0.53 and a source-destination count ($n_{sd}$) of a) 5, b) 8. The robustness parameter ($\lambda$) models end-to-end fidelity inaccuracy, with $\lambda \in \{1, 8, 16\}$, corresponding to decreasing noise magnitude.}
    \label{fig:bp_robust_conf}
\end{figure*}

\begin{figure}[tb]
\includegraphics[width=\columnwidth]{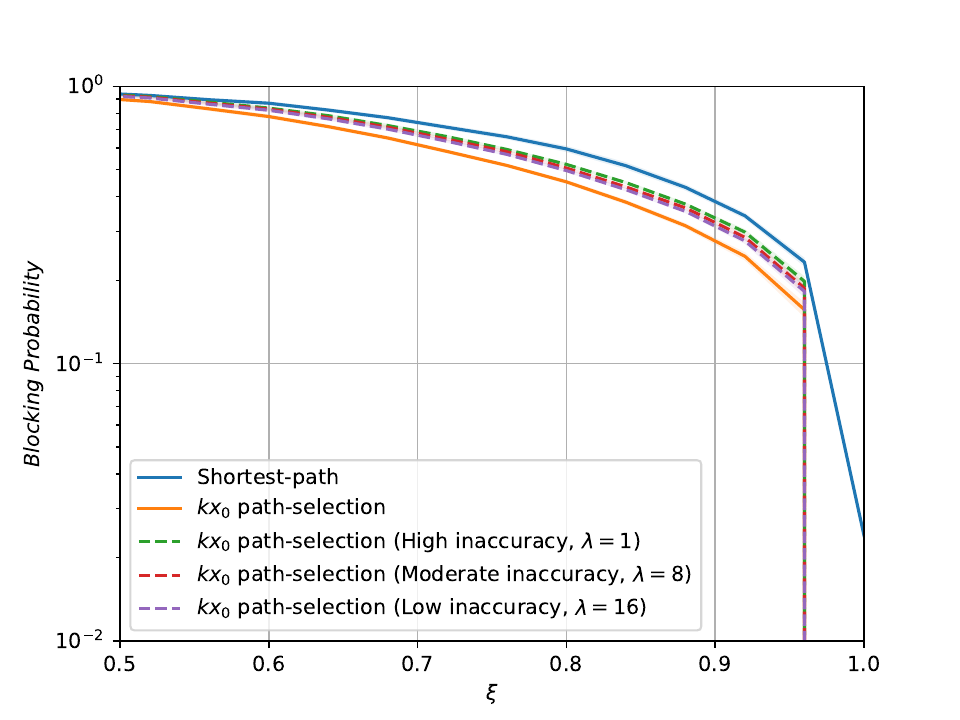}
    \vspace{-0.6cm}
    \caption{Robustness test: Blocking probability (BP) as a function of the fraction of high-quality nodes ($\xi$) in a regular transport network, with a fidelity threshold ($F_{th}$) of 0.53 and a source-destination count ($n_{sd} = 5$). The robustness parameter ($\lambda$) models end-to-end fidelity inaccuracy, with $\lambda \in \{1, 8, 16\}$, corresponding to decreasing noise magnitude.}
    \label{fig:bp_robust_conf_regular}
\end{figure}
\subsection{Robustness}
As described in Section~\ref{ssec:estimations}, the end-to-end fidelity data acquired through sampling is central to the $kx_{0}$ path-selection approach. In this aspect, we test the robustness of $kx_{0}$ approach by supplying \emph{inaccurate} end-to-end fidelity data to the algorithm. Specifically, we examine its performance on a random topology with $n_{sd} = \{ 5, 8\}$ and a regular topology with $n_{sd} = 5$. We sample the end-to-end fidelity inaccuracy from a normal distribution given by: \newline
\begin{equation}\label{eq:distribution}
Q = N \left( 0, \frac{|F_1 - F_2|}{\lambda}\right),
\end{equation}
where $F_1$ is the correct end-to-end fidelity of the path; $F_2$ is the end-to-end fidelity of the path by replacing a HQ repeater with a LQ one; $\lambda$ is a parameter modelling the end-to-end fidelity inaccuracy. Specifically, by varying $\lambda$, we modify the standard deviation of the noise added to the correct fidelity, thus controlling its magnitude. We classify these as follows: $\lambda = 1$ (High inaccuracy), $\lambda = 8$ (Moderate inaccuracy), and $\lambda = 16$ (Low inaccuracy).

Figure~\ref{fig:bp_robust_conf}a for $n_{sd} = 5$ and ~\ref{fig:bp_robust_conf}b for $n_{sd} = 8$ presents the outcomes of a robustness assessment for the $kx_{0}$ approach for $\lambda \in \{ 1, 8, 16\}$ in a random network. The blocking probability escalates as inaccuracies in the end-to-end fidelity data provided to the $kx_{0}$ approach increase or $\lambda$ decrease. For $n_{sd} = 5$, even introduction of high inaccuracy, the performance of $kx_{0}$ remains superior to that of the SP approach. The $kx_{0}$ continue to better SP under demanding condition i.e., with more SD pairs ($n_{sd} = 8$) in low and moderate inaccuracies cases. However, under more demanding conditions i.e., when fidelity inaccuracies are high along with a higher number of source-destination pairs ($n_{sd} = 8$), the advantage of $kx_0$ over SP diminishes, as shown in Figure~\ref{fig:bp_robust_conf}b. Similarly, Figure~\ref{fig:bp_robust_conf_regular} shows the robustness results for the $kx_0$ approach in a regular network for $\lambda = \{ 1, 8, 16\}$. Consistent with the random topology results, $kx_0$ outperforms SP approach even at high fidelity inaccuracies, although the performance differences between accuracy levels decrease. This reduced performance variation is due to the regular network structure, which constrains path diversity between SD pairs as compared to the random network.

Overall, these results suggest that the $kx_{0}$ approach demonstrates robust performance with inaccurate fidelity estimates, except under highly demanding conditions.

\begin{figure*}
\centering
    \begin{subfigure}{0.52\textwidth}
        \includegraphics[width=\linewidth]{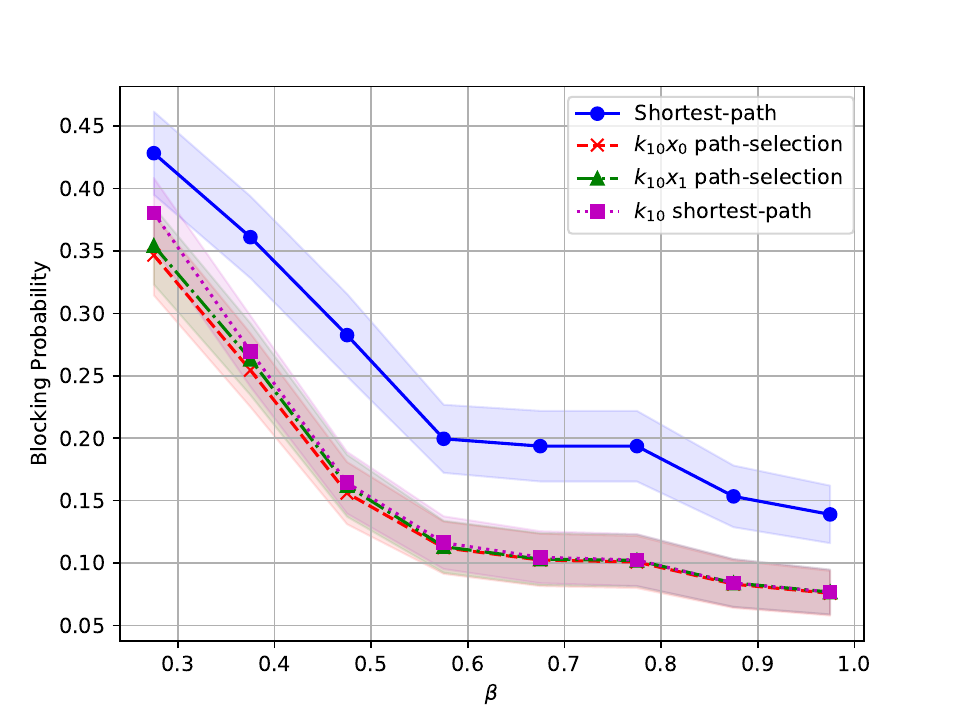}
        \caption{For $k = 10$}
        \label{fig:bp_wax_spec_conf10}
    \end{subfigure}
    \hspace{-1cm}
    \begin{subfigure}{0.52\textwidth}
        \includegraphics[width=\linewidth]{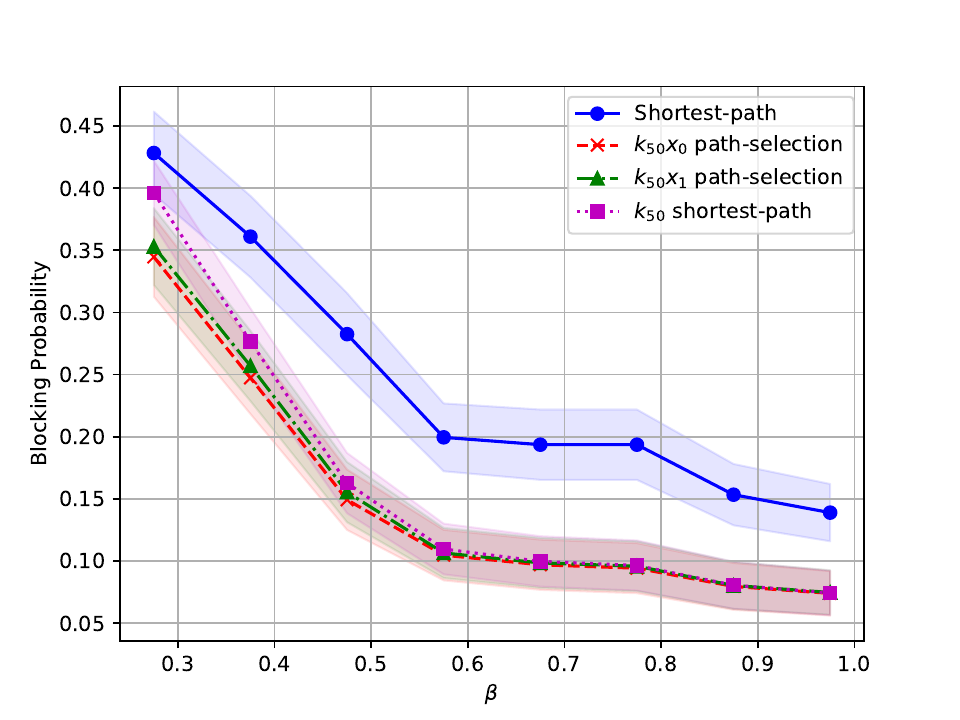}
        \caption{For $k = 50$}
        \label{fig:bp_wax_spec_conf50}
    \end{subfigure}

    \caption{Blocking probability vs. Average degree ($\beta$) with a fidelity threshold ($F_{th}$) = 0.53, number of source and destination ($n_{sd}$) = 5 in a random topology, and having range of efficiency figures of quantum repeaters. Comparison of shortest-path with grey-box algorithms having algorithm configuration parameter: a) $k = 10$, b) $k = 50$.}
    \label{fig:bp_wax_spec_conf}
\end{figure*}

\subsection{Range of Efficiency Figures}
As a final aspect, we test the performance of the considered routing algorithms by relaxing the constraint on two node classes and assigning instead efficiency figures that are randomly drawn in the range given by eq~\ref{eq:range}, as detailed in Section~\ref{sec:sim_and_method}. We set $a = 10$ to attain practically useful performances of the routing policies. Setting a higher value of parameter $a$ would decrease the BP, even achieving BP~$=0$ for higher $\beta$ values. Conversely, setting a lower value would increase the blocking probabilities above 0.5, which can be assumed to be unrealistic for practical purposes. The network configuration under consideration employs a random topology with $n_{sd} = 5$. In both the $k$-shortest path (with $k = 10$ as $k_{10}$) and $kx$-path selection (with $k = 10, \ x = 0$ as $k_{10}x_{0}$; $k = 10, \ x = 1$ as $k_{10}x_1$; $k = 50, \ x = 0$ as $k_{50}x_{0}$; $k = 50, \ x = 1$ as $k_{50}x_1$) approaches, we test the performance for $k = 10$ as well as $k = 50$. 

Figure~\ref{fig:bp_wax_spec_conf} illustrates the correlation between the blocking probability and the Waxman parameter ($\beta$), with algorithm configurations i.e., $k = 10$ (Figure~\ref{fig:bp_wax_spec_conf}a) and $k = 50$ (Figure~\ref{fig:bp_wax_spec_conf}b) comprising nodes of varying efficiency. In comparison to the SP, all the other policies yielded superior results, with $kx_{0}$ achieving top performance, as already observed in the case of two-class quality repeaters. 

With a lower $\beta$ value (i.e., lower number of links), some differences emerge, as explained in the following. The parameter $k$, that is the number of candidate shortest paths, has a minor influence on the performance, as illustrated in Figure~\ref{fig:bp_wax_spec_conf}. For a higher value i.e., $k = 50$, the performance of $kx_0$ and $kx_1$ is marginally better compared to a lower value, i.e., $k = 10$. In contrast, the performance of KSP slightly deteriorates when $k=50$. Although having additional shortest paths in the sample from which paths are selected does not seem beneficial for KSP. $kx_0$ and $kx_1$ benefit from the increased number of paths, even though the improvement is marginal. The reason is that KSP suffers from the same issue already discussed: the presence of additional paths leads to the selection of longer path lengths, which consumes valuable resources. Conversely, $kx_0$ efficiently utilises these additional paths without suffering from this drawback. 

However, as $\beta$ increases, leading to a richer diversity of potential paths, the differences in path selection among the methodologies diminish, and their performance begins to converge irrespective of parameter $k$. The greater path diversity, in fact, causes all methodologies to eventually choose the same paths, regardless of the specific parameter being optimised.


\section{Conclusions and future works}\label{sec:conclusions}
In this work, we introduced a grey-box approach to routing in quantum networks, lifting the conventional assumption that necessitates complete transparency and certainty regarding the characteristics of network components. This approach employs a sampling process to estimate the end-to-end fidelity of potential paths for source-destination pairs. Two new policies have been defined, which both assign paths on a first-come-first-serve basis: $k$ shortest-path (KSP), which assigns potentially longer/worse paths, provided that the minimum fidelity is met, to favour requests that may arrive later; and $kx$ path selection (evaluated with $x=0$ and $x=1$), which has a similar concern but only considers paths that are not $x$ hops longer than the minimum one. We have compared these policies with two alternatives via extensive simulations: shortest-path, which also does not require additional information to the network topology, and a knowledge-aware (KA) path selection that is assumed to have exact and precise knowledge about the noise figures of the quantum repeaters of the network.

A comprehensive comparison between regular and random topologies, equipped with the same resources, indicates that the performance of a routing algorithm in one topology does not necessarily translate to another. Performance is primarily influenced by the probability mass function of path lengths within each topology. For example, the performance of KA is better than that of KSP with a random topology while the performance of KSP is better than that of KA with a regular topology. The proposed $kx_0$ path-selection approach surpasses all the other approaches evaluated in delivering end-to-end fidelity paths that exceed the fidelity threshold across various topologies and number of source-destination pairs employed, while maintaining a significant fairness among users. Robustness tests confirm that the $kx_0$ path-selection method is resilient to inaccuracies in the supplied end-to-end fidelity data, underscoring its practical applicability in real-world scenarios. From our simulations, it is evident that the blocking probability per edge associated with the $kx_0$ path-selection approach scales in a manner slightly less than quadratic, showcasing its potential scalability. Additionally, this approach demonstrates superior performance in scenarios involving a diverse range of node efficiency figures.

Despite the extensive analysis in this work, some research challenges remain open for possible future investigation. First, throughput, which depends on the stochastic failure/success of local link entanglement generation and entanglement swapping, could be studied together with the fidelity of end-to-end entanglements, which is the focus of this work. Second, a system-level simulation could be performed, not limited to the path selection but also includes all the phases foreseen in a time-slot. Finally, it is possible to add priorities to the connections to provide users with a differentiated quality of service by appropriately considering them as part of the path selection algorithm.







\begin{thebibliography}{00}
\bibitem{Bennett14} Bennett, Charles H., and Gilles Brassard. "Quantum cryptography: Public key distribution and coin tossing." Theoretical computer science 560 (2014): 7-11.
\bibitem{aspuru05} Aspuru-Guzik, Alán, et al. "Simulated quantum computation of molecular energies." Science 309.5741 (2005): 1704-1707.
\bibitem{cao19} Cao, Yudong, et al. "Quantum chemistry in the age of quantum computing." Chemical reviews 119.19 (2019): 10856-10915.
\bibitem{farhi14} Farhi, Edward, Jeffrey Goldstone, and Sam Gutmann. "A quantum approximate optimization algorithm." arXiv preprint arXiv:1411.4028 (2014).
\bibitem{biamonte17} Biamonte, J., Wittek, P., Pancotti, N., Rebentrost, P., Wiebe, N., \& Lloyd, S. (2017). Quantum machine learning. Nature, 549(7671), 195-202.
\bibitem{georgescu14} Georgescu, Iulia M., Sahel Ashhab, and Franco Nori. "Quantum simulation." Reviews of Modern Physics 86, no. 1 (2014): 153.
\bibitem{shor94} Shor, Peter W. "Algorithms for quantum computation: discrete logarithms and factoring." In Proceedings 35th annual symposium on foundations of computer science, pp. 124-134. Ieee, 1994.
\bibitem{kimble08} Kimble, H. Jeff. "The quantum internet." Nature 453.7198 (2008): 1023-1030.
\bibitem{elliott02} Elliott, C. (2002). Building the quantum network. New Journal of Physics, 4(1), 46.
\bibitem{wehner18} Wehner, Stephanie, David Elkouss, and Ronald Hanson. "Quantum internet: A vision for the road ahead." Science 362, no. 6412 (2018): eaam9288.
\bibitem{broadbent09} Broadbent, Anne, Joseph Fitzsimons, and Elham Kashefi. "Universal blind quantum computation." 2009 50th annual IEEE symposium on foundations of computer science. IEEE, 2009.
\bibitem{cacciapuoti19} Cacciapuoti, Angela Sara, et al. "Quantum internet: Networking challenges in distributed quantum computing." IEEE Network 34.1 (2019): 137-143.
\bibitem{caleffi18} Caleffi, Marcello, Angela Sara Cacciapuoti, and Giuseppe Bianchi. "Quantum internet: From communication to distributed computing!." Proceedings of the 5th ACM international conference on nanoscale computing and communication. 2018.
\bibitem{ekert91} Ekert, Artur K. "Quantum cryptography based on Bell’s theorem." Physical review letters 67.6 (1991): 661.
\bibitem{pirandola20} Pirandola, Stefano, et al. "Advances in quantum cryptography." Advances in optics and photonics 12.4 (2020): 1012-1236.
\bibitem{singal22} Singal, Ansh, et al. "Hardware routed quantum key distribution networks." IET Quantum Communication 3.2 (2022): 127-138.
\bibitem{komar14} Komar, P., Kessler, E. M., Bishof, M., Jiang, L., Sørensen, A. S., Ye, J., \& Lukin, M. D. (2014). A quantum network of clocks. Nature Physics, 10(8), 582-587.
\bibitem{briegel98} Briegel, H. J., Dür, W., Cirac, J. I., \& Zoller, P. (1998). Quantum repeaters: the role of imperfect local operations in quantum communication. Physical Review Letters, 81(26), 5932.
\bibitem{duan01} Duan, L. M., Lukin, M. D., Cirac, J. I., \& Zoller, P. (2001). Long-distance quantum communication with atomic ensembles and linear optics. Nature, 414(6862), 413-418.
\bibitem{childress06} Childress, Lilian, et al. "Fault-tolerant quantum communication based on solid-state photon emitters." Physical review letters 96.7 (2006): 070504.
\bibitem{simon07} Simon, Christoph, et al. "Quantum repeaters with photon pair sources and multimode memories." Physical review letters 98.19 (2007): 190503.
\bibitem{zhao07} Zhao, Bo, et al. "Robust creation of entanglement between remote memory qubits." Physical review letters 98.24 (2007): 240502.
\bibitem{sangouard09} Sangouard, Nicolas, Romain Dubessy, and Christoph Simon. "Quantum repeaters based on single trapped ions." Physical Review A 79.4 (2009): 042340.
\bibitem{sangouard11} Sangouard, Nicolas, et al. "Quantum repeaters based on atomic ensembles and linear optics." Reviews of Modern Physics 83.1 (2011): 33.
\bibitem{li16} Li, Tao, Guo-Jian Yang, and Fu-Guo Deng. "Heralded quantum repeater for a quantum communication network based on quantum dots embedded in optical microcavities." Physical Review A 93.1 (2016): 012302.
\bibitem{Perseguers08} Perseguers, Sebastien, et al. "One-shot entanglement generation over long distances in noisy quantum networks." Physical Review A 78.6 (2008): 062324.
\bibitem{jiang09} Jiang, Liang, et al. "Quantum repeater with encoding." Physical Review A 79.3 (2009): 032325.
\bibitem{munro12} Munro, William J., et al. "Quantum communication without the necessity of quantum memories." Nature Photonics 6.11 (2012): 777-781.
\bibitem{muralidharan14} Muralidharan, Sreraman, et al. "Ultrafast and fault-tolerant quantum communication across long distances." Physical review letters 112.25 (2014): 250501.
\bibitem{munro15} Munro, William J., et al. "Inside quantum repeaters." IEEE Journal of Selected Topics in Quantum Electronics 21.3 (2015): 78-90.
\bibitem{muralidharan16} S. Muralidharan, L. Li, J. Kim, N. Lütkenhaus, M. D. Lukin, and L. Jiang, “Optimal architectures for long distance quantum communication,” Sci. Rep., vol. 6, no. 1, p. 20463, Apr. 2016.
\bibitem{illiano22} Illiano, Jessica, et al. "Quantum internet protocol stack: A comprehensive survey." Computer Networks 213 (2022): 109092.
\bibitem{yuan24} Li, Yuan, et al. "A Survey of Quantum Internet Protocols From a Layered Perspective." IEEE Communications Surveys \& Tutorials (2024).
\bibitem{zhang22} Zhang, Ling, and Qin Liu. "Optimisation of the routing protocol for quantum wireless Ad Hoc network." IET Quantum Communication 3.1 (2022): 5-12.
\bibitem{helsen23} J. Helsen and S. Wehner, ``A benchmarking procedure for quantum networks,'' npj Quantum Information, vol. 9, no. 1, p. 17, 2023.
\bibitem{meter13} Van Meter, Rodney, et al. "Path selection for quantum repeater networks." Networking Science 3 (2013): 82-95.
\bibitem{caleffi17} M. Caleffi, "Optimal Routing for Quantum Networks," in IEEE Access, vol. 5, pp. 22299-22312, 2017, doi: 10.1109/ACCESS.2017.2763325.
\bibitem{pant19} Pant, M., Krovi, H., Towsley, D. et al. Routing entanglement in the quantum internet. npj Quantum Inf 5, 25 (2019). https://doi.org/10.1038/s41534-019-0139-x
\bibitem{li21} Li, C., Li, T., Liu, YX. et al. Effective routing design for remote entanglement generation on quantum networks. npj Quantum Inf 7, 10 (2021). https://doi.org/10.1038/s41534-020-00344-4
\bibitem{cicconetti21} C. Cicconetti, M. Conti, and A. Passarella, “Request scheduling in quantum networks,” IEEE Trans. Quantum Eng., vol. 2, pp. 2–17, 2021.
\bibitem{yang22} L. Yang, Y. Zhao, H. Xu, and C. Qiao, “Online entanglement routing in quantum networks,” in 2022 IEEE/ACM 30th International Symposium on Quality of Service (IWQoS), 2022, pp. 1–10.
\bibitem{rabbie22} J. Rabbie, K. Chakraborty, G. Avis, and S. Wehner, “Designing quantum networks using preexisting infrastructure,” npj Quantum Inf., vol. 8, no. 1, p. 5, Dec. 2022.
\bibitem{li22} J. Li et al., “Fidelity-guaranteed entanglement routing in quantum networks,” IEEE Trans. Commun., pp. 1–1, 2022.
\bibitem{shi20} Shi, S., \& Qian, C. (2020). Concurrent Entanglement Routing for Quantum Networks: Model and Designs. Proceedings of the Annual conference of the ACM Special Interest Group on Data Communication on the applications, technologies, architectures, and protocols for computer communication.
\bibitem{zhang21} Shengyu Zhang, Shouqian Shi, Chen Qian, and Kwan L. Yeung, "Fragmentation-Aware Entanglement Routing for Quantum Networks," J. Lightwave Technol. 39, 4584-4591 (2021)
\bibitem{zhao21} Y. Zhao and C. Qiao, "Redundant Entanglement Provisioning and Selection for Throughput Maximization in Quantum Networks," IEEE INFOCOM 2021 - IEEE Conference on Computer Communications, Vancouver, BC, Canada, 2021, pp. 1-10, doi: 10.1109/INFOCOM42981.2021.9488850.
\bibitem{hang21} Hang Li, Jian-Peng Dou, Xiao-Ling Pang, Tian-Huai Yang, Chao-Ni Zhang, Yuan Chen, Jia-Ming Li, Ian A. Walmsley, and Xian-Min Jin, "Heralding quantum entanglement between two room-temperature atomic ensembles," Optica 8, 925-929 (2021)
\bibitem{chen24} L. Chen and Z. Jia, "On Optimum Entanglement Purification Scheduling in Quantum Networks," in IEEE Journal on Selected Areas in Communications, vol. 42, no. 7, pp. 1779-1792, July 2024, doi: 10.1109/JSAC.2024.3380080.
\bibitem{pompili21} M. Pompili et al. ,Realization of a multinode quantum network of remote solid-state qubits. Science372,259264(2021). DOI:10.1126/science.abg1919
\bibitem{axel19} Dahlberg, Axel, et al. "A link layer protocol for quantum networks." Proceedings of the ACM special interest group on data communication. 2019. 159-173.
\bibitem{helstrom69} Helstrom, Carl W. ”Quantum detection and estimation theory.” Journal
of Statistical Physics 1 (1969): 231-252
\bibitem{dur99} Dür, W., Briegel, H. J., Cirac, J. I., \& Zoller, P. (1999). Quantum repeaters based on entanglement purification. Physical Review A, 59(1), 169
\bibitem{Liu24} Liu, Maoli, et al. "Quantum BGP with Online Path Selection via Network Benchmarking." IEEE INFOCOM. 2024.
\bibitem{jli22} J. Li, Q. Jia, K. Xue, D. S. L. Wei and N. Yu, "A Connection-Oriented Entanglement Distribution Design in Quantum Networks," in IEEE Transactions on Quantum Engineering, vol. 3, pp. 1-13, 2022, Art no. 4100513, doi: 10.1109/TQE.2022.3176375.
\bibitem{humphreys18} Humphreys, P.C., Kalb, N., Morits, J.P.J. et al. Deterministic delivery of remote entanglement on a quantum network. Nature 558, 268–273 (2018). https://doi.org/10.1038/s41586-018-0200-5
\bibitem{vk23} Kumar, Vinay, et al. "Routing in Quantum Repeater Networks with Mixed Noise Figures." arXiv preprint arXiv:2310.08990 (2023).
\bibitem{wax88} Waxman, Bernard M. "Routing of multipoint connections." IEEE journal on selected areas in communications 6.9 (1988): 1617-1622.
\bibitem{skrzypczyk21} M. Skrzypczyk, S. Wehner. "An Architecture for Meeting Quality-of-Service Requirements in Multi-User Quantum Networks". arXiv preprint arXiv:2111.13124 (2021).
\bibitem{jain84} Jain, R.; Chiu, D.M.; Hawe, W.. "A Quantitative Measure of Fairness and Discrimination for Resource Allocation in Shared Computer Systems". DEC Research Report TR-301, 1984.




\end{thebibliography}
\end{document}